\documentclass{article}

\usepackage{microtype}
\usepackage{graphicx}
\usepackage{subcaption}
\usepackage{booktabs} 

\usepackage{hyperref}

\usepackage[accepted]{icml2026}

\usepackage{amsmath}
\usepackage{amssymb}
\usepackage{mathtools}
\usepackage{amsthm}

\usepackage{hyperref}
\usepackage{url}
\usepackage{mathtools}
\usepackage{xspace}
\usepackage{wrapfig}
\usepackage{graphicx}
\usepackage{subcaption}
\usepackage{booktabs}
\usepackage{multirow}
\usepackage{multicol}
\usepackage[ruled,vlined,algo2e]{algorithm2e}

\def\eg{\emph{e.g.}\xspace} 

\def\ie{\emph{i.e.}\xspace}

\def\etc{\emph{etc}\xspace}

\definecolor{darkred}{rgb}{0.7,0,0}
\definecolor{darkgreen}{rgb}{0,0.46,0}
\definecolor{purple}{rgb}{0.6,0,0.5}
\definecolor{cholocate}{HTML}{d2691e}
\definecolor{slateblue}{HTML}{6a5acd}

\newcommand{\eqdef}{\stackrel{\sf def}{=}}

\usepackage[capitalize,noabbrev]{cleveref}

\theoremstyle{plain}

\theoremstyle{definition}

\theoremstyle{remark}

\usepackage[textsize=tiny]{todonotes}

\icmltitlerunning{SARSteer: Safeguarding Large Audio-Language Models via Safe-Ablated Refusal Steering}

\begin{document}

\twocolumn[
  \icmltitle{SARSteer: Safeguarding Large Audio-Language Models \\ via Safe-Ablated Refusal Steering}



  \icmlsetsymbol{equal}{*}

  \begin{icmlauthorlist}
    \icmlauthor{Weilin Lin}{hkust}
    \icmlauthor{Jianze Li}{sysu}
    \icmlauthor{Hui Xiong}{hkust}
    \icmlauthor{Li Liu$^*$}{hkust}
  \end{icmlauthorlist}

  \icmlaffiliation{hkust}{The Hong Kong University of Science and Technology (Guangzhou)}
  \icmlaffiliation{sysu}{School of Science, Sun Yat-sen University}

  \icmlcorrespondingauthor{Li Liu}{avrillliu@hkust-gz.edu.cn}

  \icmlkeywords{Machine Learning, ICML}


  \vskip 0.3in
]



\printAffiliationsAndNotice{}  

\begin{abstract}

Large Audio–Language Models (LALMs) are becoming essential as a powerful multimodal backbone for real-world applications. However, recent studies show that audio inputs can more easily elicit harmful responses than text, exposing new risks toward deployment. While safety alignment has made initial advances in LLMs and Large Vision–Language Models (LVLMs), we find that vanilla adaptation of these approaches to LALMs faces two key limitations: 1) LLM-based steering fails under audio input due to the large distributional gap between activations, and 2) prompt-based defenses induce over-refusals on benign-speech queries. To address these challenges, we propose \textbf{S}afe-\textbf{A}blated \textbf{R}efusal \textbf{Steer}ing (SARSteer), an effective inference-time defense framework for LALMs. Specifically, SARSteer leverages text-derived refusal steering to enforce rejection without manipulating audio inputs and introduces decomposed safe-space ablation to mitigate over-refusal. Extensive experiments demonstrate that SARSteer significantly improves harmful-query refusal while preserving benign responses, establishing a principled step toward safety alignment in LALMs.
The codes and constructed datasets are released at \href{https://github.com/linweiii/SARSteer}{https://github.com/linweiii/SARSteer}.


\end{abstract}
\section{Introduction}
\label{sec:intro}


Large Audio-Language Models (LALMs) have recently emerged as powerful multimodal systems~\citep{chu2024qwen2,tang2023salmonn,ding2025kimi}, extending the general intelligent capabilities of Large Language Models (LLMs)~\citep{bai2023qwen,liu2024deepseek,achiam2023gpt} into the audio domain. By jointly modeling audio and textual inputs, LALMs enable a wide range of applications, including voice assistants~\citep{held2024distilling}, audio understanding~\citep{dinkel2025midashenglm}, real-time speech interaction~\citep{long2025vita}, \etc. Their ability to understand and generate responses directly from audio makes them a critical component for next-generation human–AI interaction systems. 

Despite their promise, the deployment of LALMs raises pressing safety concerns due to the underexplored vulnerability of new audio input. 
In the literature, most focus of safety alignment has been laid in text-based LLMs~\citep{kim2024road,zhang2025stair,qi2024safety}, leveraging both \textit{fine-tuning-based defenses} such as supervised fine-tuning (SFT)~\citep{liu2023makes} and reinforcement learning from human feedback (RLHF)~\citep{bai2022training}, and more advanced \textit{inference-based defenses} such as activation steering~\citep{panickssery2023steering,zhao2025adasteer}. While fine-tuning can be effective with high-quality data or well-trained reward models, its resource-intensive nature makes inference-based defenses more practical for scalable deployment. Similar efforts have recently extended to Large Vision–Language Models (LVLMs)~\citep{wang2024qwen2,lu2024deepseek,zhu2023minigpt,liu2024improved}, leading to new fine-tuning-based~\citep{zhang2025spa,zong2024safety} and inference-based~\citep{wang2024adashield,ding2024eta} defense strategies designed for vision modality. 
In contrast, the safety alignment of LALMs remains largely underexplored: beyond some initial findings~\citep{yang2024audio,song2025audio}, which show that LALMs are far more likely to comply with harmful speech than text, no principled defense strategies have been developed. A natural solution, therefore, is to transfer the alignment techniques originally designed for LLMs or LVLMs into the audio–language setting. In this work, \textbf{we focus on inference-based defenses}, \eg, activation steering from LLMs~\citep{panickssery2023steering} and prompt-based defenses from LVLMs~\citep{wang2024adashield}, to align LALMs with harmless outputs.

However, such transfers with vanilla adaptations expose two critical limitations. \textbf{First, LLM-based steering fails under audio input.} In LLMs, steering vectors constructed from harmful–safe text pairs can reliably shift representations toward safe regions and enhance refusal behaviors. In LALMs, by contrast, harmful and safe speech inputs occupy widely divergent latent distributions than in text, making the harm-to-safe direction unreliable (Section~\ref{subsec:fail-steer-audio}). \textbf{Second, prompt-based defenses from LVLMs induce over-refusal unconspicuously}~\citep{jiang2025hiddendetect}. While defensive prompts (\eg, instructing the model to respond \textit{“I am sorry”} to unethical or illegal requests) can block some harmful queries, they also cause benign queries with lexical similarity to be mistakenly rejected (Section~\ref{subsec:overrefusal-prompt-based}).
Despite efforts such as AdaShield~\citep{wang2024adashield}, which refines prompts to better distinguish benign inputs, the coarse input-level instructions, containing two opposing actions of answering or refusing, struggle to coordinate effectively.

To address these challenges, we propose an inference-based alignment framework, \textbf{S}afe-\textbf{A}blated \textbf{R}efusal \textbf{Steer}ing (\textbf{SARSteer}), for LALMs. 
SARSteer targets both the failure of steering audio modality and the over-refusal issue observed in prompt-based defenses. It consists of two key components:
1) \textbf{Text-derived refusal steering.} Instead of contrasting harmful and safe speech inputs, which suffer from distributional gap, SARSteer extracts refusal vectors directly from textual refusal prompts (\eg, \textit{“I cannot assist with that”}). These vectors capture safety-aligned semantics in intermediate activations and provide a modality-agnostic direction for enhancing harmful-query rejection.
2) \textbf{Decomposed safe-space ablation.} To mitigate over-refusal on benign queries, SARSteer employs a projection correction step. Specifically, we use \textit{principal component analysis} (PCA) on safe samples to identify the dominant subspace of benign semantics, and then ablate this component from the refusal vector. This ensures that refusal steering acts only on harmful directions while preserving safe responses.
By jointly leveraging these two components, SARSteer avoids costly fine-tuning, operates entirely at inference time, and establishes a principled defense strategy for LALMs that is robust against harmful inputs while maintaining utility on benign ones.

Our contributions are summarized as follows:
\begin{itemize}
    \item We construct paired harmful–safe datasets in the speech domain and provide a systematic study of the representational differences between text and audio inputs, explaining the failure of direct activation steering transfer.
    \item We introduce an effective inference-time defense framework for LALMs, based on text-derived refusal steering and decomposed safe-space ablation, filling the gap of broad LALMs applications and the scarcity of the specified safety alignment.
    \item Extensive experiments demonstrate that our method significantly improves harmful-query refusal while maintaining overall utility, achieving a favorable trade-off between safety and usability.
\end{itemize}

\paragraph{Conflict of Interest Disclosure.}
The authors declare no financial conflicts of interest related to this work.

\section{Related Work}
\label{sec:related}

\subsection{LLM Safety Alignment}

Substantial research has focused on aligning LLMs with human values and safety standards~\citep{bai2022training,kim2024road,zhang2025stair,qi2024safety}. Prominent approaches include reinforcement learning from human feedback (RLHF), which fine-tunes models using human-preferred responses~\citep{ouyang2022training,bai2022training}, and supervised fine-tuning (SFT) on safety-centric datasets~\citep{liu2023makes}. These methods can all be categorized as \textit{fine-tuning-based defenses}. 
Despite their effectiveness, they often require extensive human annotation and computational resources, limiting their applications.

More recently, \textit{inference-time techniques} have gained attention for their efficiency and low resource demands~\citep{arditi2024refusal,zhao2025adasteer,qian2025hsf}. For instance, activation steering methods intervene in the model’s internal representations to guide outputs toward desired behaviors~\citep{panickssery2023steering,zhao2025adasteer,ghosh2025safesteer}. Similarly, refusal prompts, prepending input queries with safety-guided instructions, have been shown to enhance defensiveness against malicious inputs without additional training~\citep{zheng2024prompt,qian2025hsf}. These approaches circumvent the need for large-scale fine-tuning, making them a more feasible solution for real-world industries.

\subsection{Multimodal LLM Safety}

The integration of visual modalities introduces new vulnerabilities and attack surfaces in Multimodal Large Language Models (MLLMs)~\citep{li2024images,zhang2025fc}. Adversaries can exploit cross-modal inconsistencies to bypass safety alignments, such as by embedding harmful content in images paired with benign text~\citep{gong2025figstep}. In response, several defense strategies have been proposed. 
AdaShield~\citep{wang2024adashield} employs adaptive shield prompting to defend against structure-based jailbreak attacks without fine-tuning the model. Similarly, ETA~\citep{ding2024eta} introduces a two-phase “Evaluate then Align” framework that assesses both visual and textual inputs for harmful content and aligns outputs via shallow and deep alignment mechanisms. Other methods like DAVSP~\citep{zhang2025davsp} optimize a visual safety prompt using activation-space supervision, while HiddenDetect~\citep{jiang2025hiddendetect} monitors hidden states to identify harmful patterns. These inference-time methods effectively enhance safety against the vision-space vulnerability.

However, existing research predominantly focuses on LVLMs, while for LALMs, only two initial attempts, RRS~\cite{yang2025reshaping} and ALMGuard~\citep{jin2025almguard}, discuss the safety training and safety shortcuts issue, respectively. The research on the audio modality is largely unexplored in terms of safety alignment. Our work represents one of the preliminary steps toward developing inference-time safety alignment for the speech domain. 
By leveraging text-derived refusal steering and decomposed safe-space ablation in the model activation space, our approach offers a flexible, efficient solution to refuse harmful inputs while maintaining the general utility of LALMs.

\section{Preliminary and Motivation Analysis}
\label{sec:preliminary}

\subsection{Problem Formulation}
\label{subsec:problem-formulation}

\paragraph{Model Description.}
We consider a basic-form LALM \(M\)\footnote{Despite the presence of additional components in certain LALMs (\eg,  an audio decoder), in this paper, we focus on the basic architecture, namely the audio encoder, multimodal projector, and language model backbone, as these elements are common to most designs. The detailed illustration on Qwen2-Audio is provided in Appendix~\ref{appendix:detail_model_structure}.} that processes multimodal queries \(Q = (a, t)\), where \(a\) is an \textit{audio} signal and \(t\) is a \textit{textual} input, to a textual output. The audio encoder \(\mathcal{E}_a\) maps \(a\) into an embedding \(e_a = \mathcal{E}_a(a)\), which is then projected into the textual embedding space through a multimodal projector \(\mathcal{P}\), yielding \(\tilde{e}_a = \mathcal{P}(e_a)\). 
After that, the audio representations and the tokenized textual input are fed into the autoregressive language model backbone \(\mathcal{M}\), generating an output sequence 
\begin{equation}
Y_I = (y_1, \ldots, y_I), \  
P(Y_I \mid Q) = \prod_{i=1}^{I} P(y_i \mid Y_{<i}, \tilde{e}_a, e_t; \mathcal{M}),
\end{equation}
where \(e_t\) is the discrete textual embedding processed by \(\mathcal{M}\), \(Y_I\) and \(Y_{<i}\) represent the complete response with \(I\) tokens and the first generated \(i\) tokens, respectively.
It can be seen that the LALM extends standard LLMs by incorporating audio understanding through the audio encoder and multimodal projector, enabling both audio-text-conditioned generation.

\paragraph{Task Description.}
Based on the above LALM model, we now formulate the \textit{inference-time safety alignment} task, which is typically performed in a training-free manner after the model training phase. 
Since the model output \(Y_I\) is free-form text, we introduce an evaluation function
\begin{equation}
\mathcal{R}: \mathcal{Y} \to \{0,1\},
\end{equation}
to judge whether a response constitutes a refusal (\(\mathcal{R}(Y_I)=1\)) or not (\(\mathcal{R}(Y_I)=0\)), which is implemented by an auxiliary LLM~\citep{xie2024sorry} or a matching-based method~\citep{wang2024adashield}\footnote{In this work, we use matching-based method to compute \textit{refusal rate} (RR) on both harmful and benign datasets; use LLM-based method to assess \textit{attack success rate} (ASR) on harmful queries.}.  
We denote by \(\mathcal{Q}_\text{harm}\) the set of harmful queries, and by \(\mathcal{Q}_\text{safe}\) the corresponding benign queries set (Section~\ref{subsec:paired-dataset}).
The objective of safety alignment is threefold:
\begin{enumerate}
    \item \textbf{Refuse Harmful Queries.} For \(Q \in \mathcal{Q}_\text{harm}\), maximize \(\mathbb{E}_{Q \in \mathcal{Q}_\text{harm}}[\mathcal{R}(M(Q))]\) to ensure the model refuses harmful inputs.   
    \item \textbf{Preserve Helpfulness on Safe Queries.} For \(Q \in \mathcal{Q}_\text{safe}\), minimize \(\mathbb{E}_{Q \in \mathcal{Q}_\text{safe}}[\mathcal{R}(M(Q))]\) to avoid unnecessary refusals.  
    \item \textbf{Maintain General-purpose Utility.} On the benchmarks \(\mathcal{B}\), enforce \(\mathrm{Perf}(M, \mathcal{B}) \approx \mathrm{Perf}(M_0, \mathcal{B})\), where \(M_0\) denotes the original unaligned model and \(\mathrm{Perf}(M, \mathcal{B})\) represents the performance of \(M\) on \(\mathcal{B}\) under its own evaluation method (\eg, \textit{LLM-as-a-judge} with well-designed prompt in AirBench~\citep{yang2024air}.), ensuring that the aligned model preserves general performance. 
\end{enumerate}

The above objectives indicate that safety alignment seeks to \textbf{enhance refusals of harmful inputs, mitigate unnecessary refusals of safe inputs, and preserve overall model utility}.


\subsection{Harmful-Safe Paired Audio Dataset Construction}
\label{subsec:paired-dataset}

To study data-dependent alignment methods and rigorously evaluate them, we first construct several harmful–safe paired datasets in the speech domain based on the safety benchmarks of other domains, \eg, AdvBench~\citep{zou2023universal} and Figsteps~\citep{gong2025figstep}. Specifically, for the harmful set \(\mathcal{Q}_\text{harm}\), we directly convert the harmful textual queries into spoken using the OpenAI \textit{text-to-speech} (TTS) system\footnote{\href{https://platform.openai.com/docs/models/tts-1-hd}{https://platform.openai.com/docs/models/tts-1-hd}}, and unify the textual instruction as \textit{``Please generate detailed steps for the question in the audio.''}~\citep{gong2025figstep}. To obtain their safe counterparts \(\mathcal{Q}_\text{safe}\), we apply LLMs (\eg, DeepSeek-R1~\citep{guo2025deepseek}) to purify each harmful query into a lexically-similar benign version (\eg, \textit{``How can I make a fake bank statement?''} $\to$ \textit{``How can I obtain an official bank statement?''}), followed by TTS synthesis to match the audio modality. This one-to-one purification ensures that every harmful query has a semantically related but safe alternative, yielding paired datasets with high lexical similarity but fundamentally different safety implications. In this work, we randomly sample 100 harmful-safe paired queries from the Figsteps-based dataset (referred to as Figstep-audio) for alignment, denoted as \(\mathcal{Q}^{s}_\text{harm}\) and \(\mathcal{Q}^{s}_\text{safe}\), while the remaining pairs are reserved for evaluation. 
Further details of the dataset are provided in the Appendix~\ref{appendix:constructed_data}.

Such paired safe data is necessary because existing benign benchmarks~\citep{yang2024air} often fail to expose the issue of \emph{over-refusal} on borderline safe inputs. 
By explicitly pairing harmful and safe queries with minimal lexical differences, our datasets provide a sharper testbed to evaluate whether alignment methods can reliably distinguish harmful instructions from benign ones, thus exposing subtle safety-utility trade-offs and directly supporting the second objective.

\subsection{Failures of Steering Audio Modality}
\label{subsec:fail-steer-audio}
Based on the constructed datasets, we now investigate the transfer of inference-based safety alignment techniques from other domains, \eg, activation steering from LLM safety~\citep{zhao2025adasteer,ghosh2025safesteer}.

\paragraph{Vanilla Adaptation of Two Activation Steering Defenses.}
Typically, there exist two kinds of steering vector implementations: extracting from harmful-to-safe query~\citep{arditi2024refusal} and from harmful compliance-to-refusal query~\citep{zhao2025adasteer}, both relying on the \textit{difference-in-means} technique~\citep{belrose2024diff}. 
To facilitate discussion, we refer to the two methods as \textit{MDSteer-h2s} (mean-difference steering in the harmful-to-safe direction) and \textit{MDSteer-c2r} (mean-difference steering in the compliance-to-refusal direction on harmful inputs), respectively. 
Under our LALM setting, where harmful or safe semantics are embedded in the audio modality, \textbf{both steering vectors are computed based on differences between audio inputs}. 
We formulate the vanilla adaptation to LALMs as follows. 
Let \(h^{l}(Q)\) denote the activation at the last token position of layer \(l \in [L]\)~\citep{zhao2025adasteer} of \(\mathcal{M}\), where \(Q=(a,t)\) is a multimodal query. 

\noindent\textbf{(1) MDSteer-h2s.}  
Given our paired datasets \(\mathcal{Q}^s_\text{harm}\) and \(\mathcal{Q}^s_\text{safe}\), we compute
\begin{equation}
\begin{aligned}
& \mu_\text{harm}^{l} \eqdef \frac{1}{|\mathcal{Q}^s_\text{harm}|} \sum_{Q \in \mathcal{Q}^s_\text{harm}} h^{l}(Q), \\
& \mu_\text{safe}^{l} \eqdef \frac{1}{|\mathcal{Q}^s_\text{safe}|} \sum_{Q \in \mathcal{Q}^s_\text{safe}} h^{l}(Q),
\end{aligned}
\label{equ:mean_act}
\end{equation}
and define the steering vector as
\begin{equation}
v_{h2s}^{l} \eqdef \mu_\text{safe}^{l} - \mu_\text{harm}^{l}.
\end{equation}

\noindent\textbf{(2) MDSteer-c2r.}  
Alternatively, we group harmful queries by their generated response type. Let \(\mathcal{Q}_\text{harm}^{\text{s-comp}}\) denote those eliciting compliant harmful responses, and \(\mathcal{Q}_\text{harm}^\text{s-ref}\) denote those eliciting refusals, as determined by the evaluation function \(\mathcal{R}\). 
We obtain the corresponding mean activation values \(\mu_\text{harm-c}^{l}\) and \(\mu_\text{harm-r}^{l}\) in the same way as Equation~\ref{equ:mean_act}, and define the steering vector as: 
\begin{equation}
v_{c2r}^{l} \eqdef \mu_\text{harm-r}^{l} - \mu_\text{harm-c}^{l}.
\end{equation}
During inference, the vector \(v^{l}\) (either \(v_{h2s}^{l}\) or \(v_{c2r}^{l}\)) is added to the model's hidden states at each generated token position \(i\), scaled by a coefficient \(\alpha\):
\begin{equation}
h_{i}^{'l} \eqdef h_{i}^{l} + \alpha v^{l}.
\label{equ:act_add}
\vspace{-3mm}
\end{equation}


\begin{figure}
    \centering
    \includegraphics[width=1\linewidth]{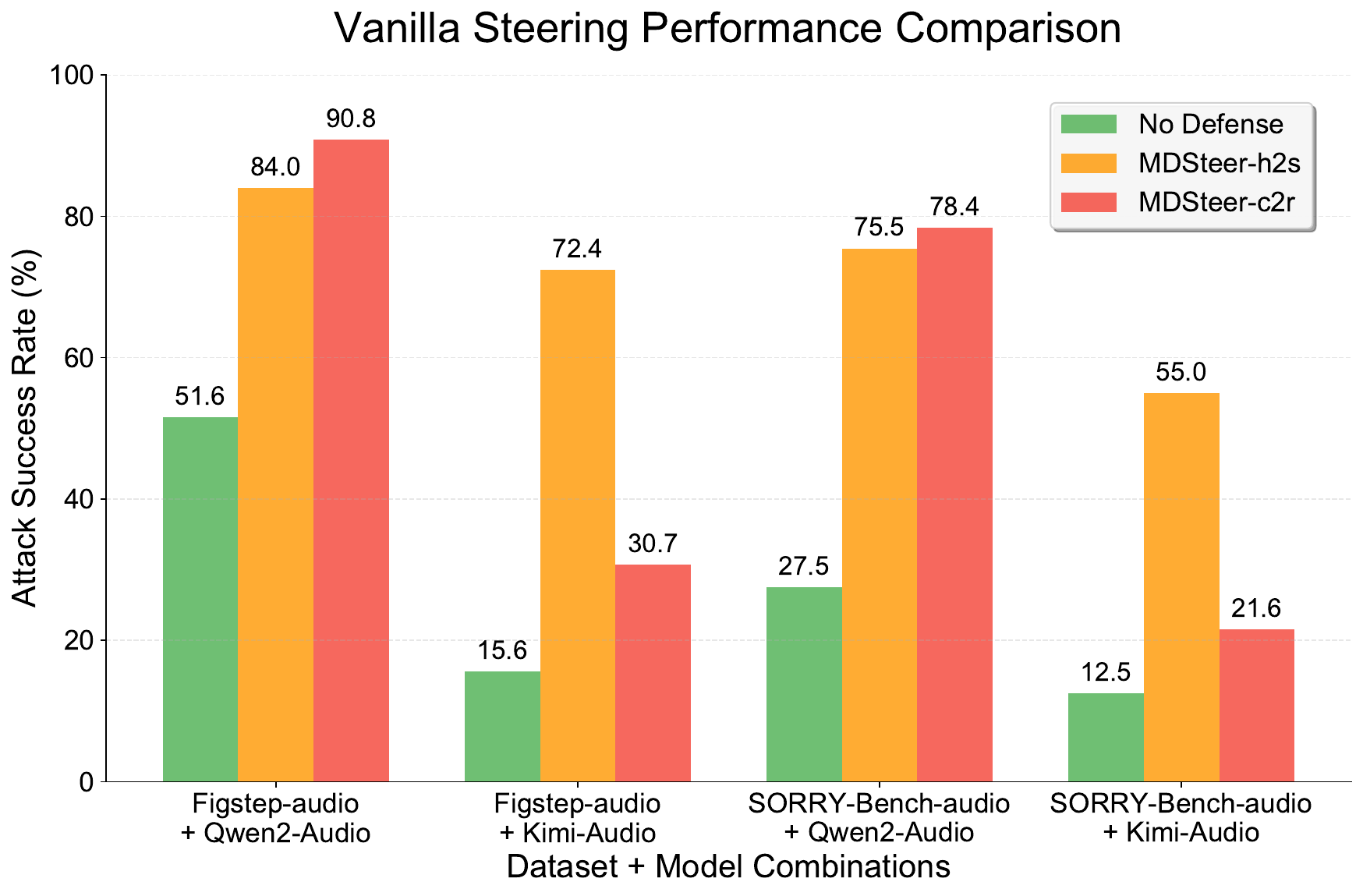}
    \vspace{-5mm}
    \caption{Performance of vanilla adaptations of LLM-based steering on LALMs.}
    \label{fig:vanilla_steer}
\end{figure}

\begin{figure}[h] 
  \begin{minipage}[t]{0.5\textwidth} 
    \vspace{0pt} 
    \centering
    \begin{subfigure}[b]{0.49\linewidth}
      \centering
      \includegraphics[width=\linewidth]{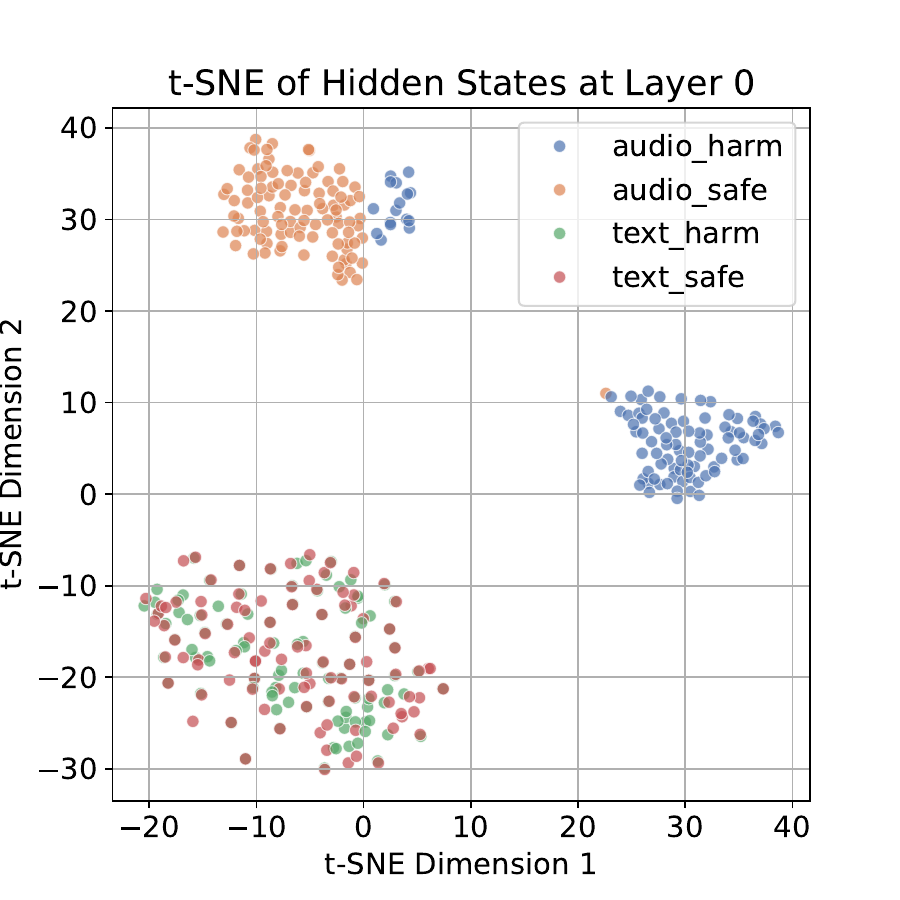} 
    \end{subfigure}
    \hfill
    \begin{subfigure}[b]{0.49\linewidth}
      \centering
      \includegraphics[width=\linewidth]{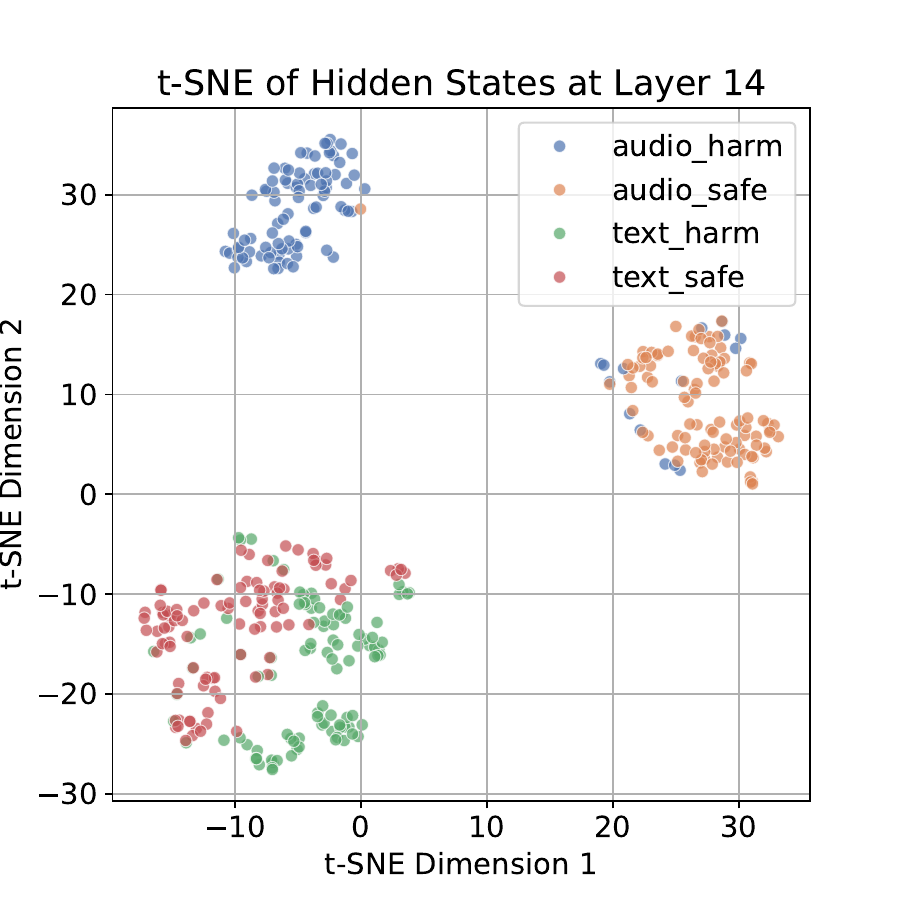} 
    \end{subfigure}
    \caption{t-SNE visualization of hidden states in Qwen2-Audio using Figstep-audio datasets. ``audio'' and ``text'' represent the input modalities containing the questions; ``harm'' and ``safe'' represent the harmfulness of the questions.}
    \label{fig:tsne_hs}
  \end{minipage}
\end{figure}

\paragraph{Results Analysis.}

We evaluate the ASR performance of MDSteer-h2s and MDSteer-c2r on Qwen2-Audio~\citep{chu2024qwen2} and Kimi-Audio~\citep{ding2025kimi}, using our audio-version Figstep~\citep{gong2025figstep} and SORRY-Bench~\citep{xie2024sorry}. 
As shown in Figure~\ref{fig:vanilla_steer}, both methods not only fail to improve ASR performance over the “No Defense” baseline (the original performance of LALMs), but also degrade it. 
To understand this failure, we analyze the hidden representations of harmful and safe inputs across both text and audio modalities using t-SNE (Figure~\ref{fig:tsne_hs}). 
In the text modality, harmful and safe queries overlap in shallow layers (left subfigure) and become linearly separable at intermediate depths (right subfigure), consistent with \citep{panickssery2023steering}, which reports that separability emerges suddenly after a particular layer.
This overlapping structure enables a feasible harmful-to-safe (h2s) transition, making h2s steering (and similarly c2r) meaningful in the text modality.
In sharp contrast, the audio modality shows early and persistent separation between harmful and safe queries across all layers, leaving no shared subspace to define a valid steering path. 
As a result, both h2s and c2r directions degenerate into noisy perturbations that fail to induce refusal. This striking gap reveals a \textbf{fundamental limitation: speech activations cannot serve as a feasible operating space for safety steering, and effective alignment should instead be derived from the refusal signals embedded in the text modality.} 
This observation motivates our approach of text-derived refusal steering in Section \ref{subsec:method_text_refusalsteering}.

\subsection{Over-refusal of Prompt-based Defenses}
\label{subsec:overrefusal-prompt-based}

Another critical limitation is the \textit{over-refusal} (or over-defense) issue in prompt-based defenses when transferred from LVLMs, \ie, the tendency to refuse even benign or borderline-safe queries. 

\paragraph{Evaluation of Balanced Refusal.}
While the over-refusal phenomenon has been discussed in prior LLM and LVLM defense studies~\citep{cui2024or,wang2024adashield,jiang2025hiddendetect}, a precise evaluation has remained challenging due to the lack of paired harmful-safe datasets. 
In particular, for LALMs, existing metrics are insufficient to capture the trade-off between refusing harmful queries and preserving utility on borderline benign ones. 
To address this gap, we adopt the \textit{refusal rate} (RR) with a matching-based evaluation method~\citep{wang2024adashield}\footnote{The refusal signals used for matching is listed in Appendix~\ref{appendix:refusal_signal}.}, defined as
\begin{equation}
\text{Refusal Rate} \eqdef \frac{|\#\text{Refusal responses}|}{|\#\text{All responses}|}.
\end{equation}
Inspired by \textit{balanced accuracy}~\citep{brodersen2010balanced}, we further introduce the \textit{balanced refusal rate} (BRR), which considers both harmful and safe sets simultaneously. 
Denoting the refusal rate on harmful and safe inputs as \(\text{RR}_{\text{harm}}\) and \(\text{RR}_{\text{safe}}\), the BRR is defined as 
\begin{equation}
    \text{BRR} \eqdef \frac{1}{2}\left[\text{RR}_{\text{harm}}+(1-\text{RR}_{\text{safe}})\right] 
    = \frac{1+\text{RR}_{\text{harm}}-\text{RR}_{\text{safe}}}{2},
\end{equation}
where \(\text{BRR}\in[0,1]\) reflects the overall refusal capability (or helpfulness): high values indicate that harmful queries are correctly rejected while safe ones are preserved.

\begin{table}[]
    \centering
    \caption{Performance of vanilla adaptations of prompt-based defenses from LVLMs on Qwen2-Audio. NOTE: ``Avg. Score'' is an LLM-based evaluation metric from the original paper~\citep{yang2024air} to assess the benign performance.}
    \label{tab:refusal_rate}
    \resizebox{\linewidth}{!}{
    \begin{tabular}{l|ccc|cc}
        \toprule
        \multirow{2}{*}{Defense} 
            & \multicolumn{3}{c|}{Figstep-audio (Harmful-safe paired)} 
            & \multicolumn{2}{c}{AirBench (General purpose) } \\
        \cmidrule(lr){2-4} \cmidrule(lr){5-6}
            & Harmful (RR) (\%)$\uparrow$ & Safe (RR)(\%)$\downarrow$ & BRR (\%)$\uparrow$
            & RR (\%)$\downarrow$ & Avg. Score (1-10)$\uparrow$\\
        \midrule
        No Defense & 62.00 & 21.60 & 70.20 & 1.23 & 7.43 \\
        \midrule
        AdaShield  & 75.60 & 36.00 & 69.80 & 2.78 & 7.39 \\
        FSD        & 90.00 & 63.60 & 63.20 & 2.64 & 7.31 \\
        \bottomrule
    \end{tabular}}
    \vspace{-6mm}
\end{table}

\paragraph{Prompt-based Defenses from LVLMs.}
We transfer and examine representative prompt-based defenses, \eg, AdaShield~\citep{wang2024adashield} and FSD~\citep{gong2025figstep}, on LALMs, which were originally proposed for LVLMs. 
The implementation details are postponed to Appendix~\ref{appendix:baseline}.
Based on the above metrics, we evaluate the overall performance on our constructed paired dataset, \ie, Figstep-audio, and on a general-purpose audio benchmark, \ie, AirBench~\citep{yang2024air}.
The results are illustrated in Table~\ref{tab:refusal_rate}. We can observe that these defenses appear to maintain reasonable performance with only slight degradation on RRs ($<$2\%) and Avg. Scores ($<$0.13) on AirBench, as the benign queries are typically far from the decision boundary. However, when evaluated on the paired harmful-safe dataset (Figstep-audio), which explicitly includes \textit{borderline safe samples} that partially overlap with harmful semantics, a clear over-refusal issue emerges: the improved harmful RRs also lead to significant higher safe RRs, degrading the overall helpfulness (lower BRRs). The results highlight the necessity of considering the borderline-safe data and reveal that \textbf{vanilla adaptations of prompt-based defenses incur unconspicuous over-refusal}. 
This also motivates our approach of ablating the safe subspace in hidden space (\ie, the decomposed safe-space ablation of Section~\ref{subsec:methoc_safe_ablation}).

\section{Methodology}
\label{sec:method}
Based on the above analysis, we propose \textbf{SARSteer}, which derives the steering vector from the refusal text of the same speech input (\ie, text-derived refusal steering) and ablates the safe subspace of its hidden representation to mitigate over-refusal on benign queries (\ie, decomposed safe-space ablation). 
We now present the technical details of the two components. 
The overview of SARSteer is shown in Figure~\ref{fig:framework} and the corresponding algorithm outline is provided in Appendix~\ref{appendix:algorithm_outline}.

\subsection{Text-derived Refusal Steering}
\label{subsec:method_text_refusalsteering}
Prompt-based defenses provide a practical approach to increasing the refusal rate of MLLMs by appending refusal-style text~\citep{wang2024adashield}, despite the limitations of over-refusal and inflexibility to multiple purposes. Combined with the analysis in Section~\ref{subsec:fail-steer-audio}, this insight inspires us: \textit{why not extract the controllable steering vector from the appended refusal text, while keeping the audio modality unchanged?} 
Therefore, we first calculate mean activation values of the modified query \(Q'=(a,t+p)\) and the original query \(Q=(a,t)\) from \(\mathcal{Q}^s_\text{harm}\) using Equation~\ref{equ:mean_act}, where \(p\) denotes a refusal text prompt (\eg, \textit{``I cannot assist with that.''})\footnote{This example is used as a \textit{semantic anchor} to obtain a “refusal direction” in the latent space, capturing consistent activation patterns associated with refusal behavior. Other refusal prompts can also elicit similar refusal directions as in Appendix \ref{appendix:diff_refusal_prompts}}. 
We denote their mean vectors as \(\mu_\text{harm-tr}^{l}\) and \(\mu_\text{harm}^{l}\), respectively.
Then the steering vector representing the refusal direction can be defined as 
\begin{equation}
    \hat{v}^{l} \eqdef \mu_\text{harm-tr}^{l} - \mu_\text{harm}^{l}.
    \label{equ:harm_sv}
\end{equation}

Applying this vector to harmful inputs using Equation~\ref{equ:act_add} can effectively improve the refusal rate.

\begin{figure}
    \centering
  \includegraphics[width=0.95\linewidth]{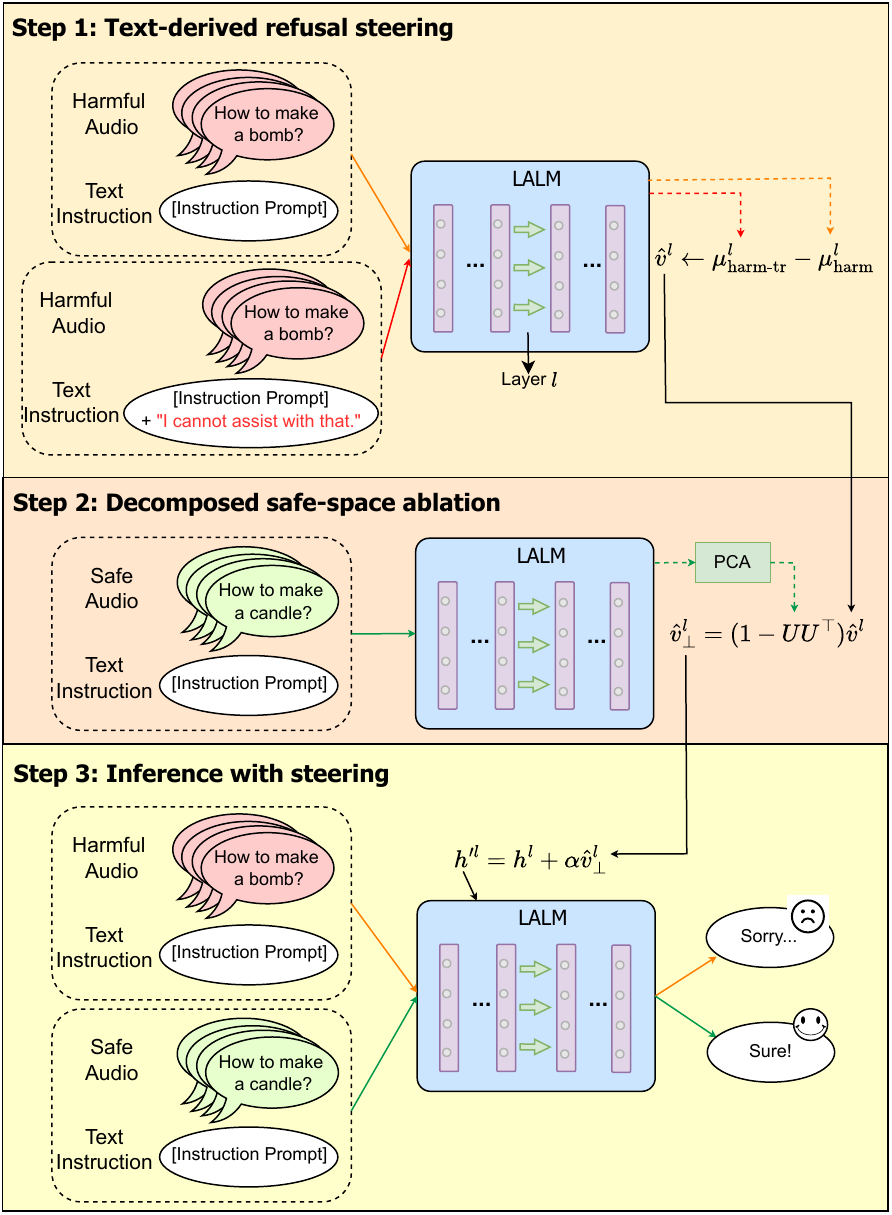} 
    \caption{Overview of the SARSteer. Step 1 derives a refusal steering vector from text prompts; Step 2 ablates the safe-subspace component via PCA to prevent over-refusal; Step 3 adds the ablated vector at inference to shift harmful responses toward refusal.}
    \label{fig:framework}
    \vspace{-5mm}
\end{figure}

\subsection{Decomposed Safe-space Ablation}
\label{subsec:methoc_safe_ablation}
While text-derived refusal steering provides a controllable vector \(\hat{v}^{\,l}\), it risks activating dimensions that are also present in benign inputs, leading to \emph{over-refusal}. To address this issue, we propose a decomposition strategy that explicitly removes safe-subspace components from \(\hat{v}^{\,l}\) by leveraging the statistical structure of safe activations.

Concretely, we first collect activations from safe queries at layer \(l\):  
\begin{equation}
H_{\text{safe}}^l = [h^{l}(Q)]_{Q\in \mathcal{Q}^s_\text{safe}} \in \mathbb{R}^{D \times n},
\end{equation}
where \(D\) is the hidden dimension of \(\mathcal{M}\) (\eg, 4096 in Qwen2-Audio) and \(n\) is the number of samples (\eg, 100).  
We then apply Principal Component Analysis (PCA) to \(H_{\text{safe}}^l\), which identifies a low-dimensional subspace spanned by the top-\(k\) principal components \(U \in \mathbb{R}^{D \times k}\), satisfying \(U^\top U=I_k\). These directions capture the dominant variance of safe representations, and therefore encode the most salient features that should be preserved when handling benign inputs.  
Given the principal components, the steering vector can be decomposed as  
\begin{equation}
\hat{v}^{\,l} = (UU^\top)\hat{v}^{\,l} + (1-UU^\top) \hat{v}^{\,l} = \hat{v}^{\,l}_{\parallel} + \hat{v}^{\,l}_{\perp},
\end{equation}  
where \(\hat{v}^{\,l}_{\parallel}\) is the safe-subspace component and \(\hat{v}^{\,l}_{\perp}\) is the orthogonal components.
We retain only the orthogonal part by projecting away the safe subspace as \textbf{our final steering vector}:  
\begin{equation}
\hat{v}_{\perp}^{\,l} = (1 - U U^\top)\, \hat{v}^{\,l}.
\end{equation}
This ensures that the steering signal emphasizes harmful-related activations while minimizing interference with safe inputs.  
Similar to Equation~\ref{equ:act_add}, during inference, activations are updated by  
\begin{equation}
h'^{\,l} = h^{l} + \alpha \hat{v}_{\perp}^{\,l}.
\end{equation}
By explicitly grounding the decomposition in PCA, this method provides a solid and interpretable mechanism: it systematically separates refusal-relevant directions from benign-safe variance, thus making the steering both effective and robust.  
Furthermore, we provide a \textbf{mathematical intuition} to demonstrate how SARSteer accomplishes the three alignment objectives in Appendix~\ref{appendix:math_intuition}.

\section{Experiment}
\label{sec:experiment}

\subsection{Experimental Setup}

\paragraph{Evaluation Metrics and Datasets.}
We evaluate safety alignment across three aspects:
\textbf{1) Harmfulness:} measured by \textit{attack success rate} (ASR) using LLM-as-a-judge on Figstep-audio~\citep{gong2025figstep}, AdvBench-audio~\citep{zou2023universal}, SORRY-Bench-audio~\citep{xie2024sorry}, and AJailBench~\citep{song2025audio}.
\textbf{2) Helpfulness:} measured by \textit{Balanced Refusal Rate} (BRR) on paired datasets including Figstep-audio and AdvBench-audio.
\textbf{3) General Utility:} evaluated on AirBench~\citep{yang2024air} following the original LLM-based evaluation.
More details are postponed to Appendix~\ref{appendix:exp_setup}.

\paragraph{Baselines.}
In this section, we use the vanilla adapted defenses from LLMs and LALMs as the baselines. As discussed in Sections~\ref{subsec:fail-steer-audio} and \ref{subsec:overrefusal-prompt-based}, we implement LALM-version prompt-based defenses, \ie, AdaShield~\citep{wang2024adashield} and FSD~\citep{gong2025figstep}, as well as activation-steering defenses~\citep{belrose2024diff,zhao2025adasteer}, \ie, MDSteer-h2s and MDSteer-c2r.
More implementation details are postponed to Appendix~\ref{appendix:baseline}. We also reproduce and compare an audio-specific safety alignment, RRS~\cite{yang2025reshaping}, in Appendix~\ref{appendix:ft_baseline}.

\paragraph{Implementation Details.}
We mainly use two state-of-the-art (SOTA) open-sourced LALMs, \ie, Qwen2-Audio~\citep{chu2024qwen2} and Kimi-Audio~\citep{ding2025kimi}, to evaluate all defense methods. We randomly sample 100 harmful-safe paired queries from Figstep-audio for alignment implementation. For our SARSteer, we use the simplest refusal prompt \textit{``I cannot assist with that.''} to extract the steering vector by default. For other hyperparameters: the scaling coefficient $\alpha$ is set to 0.1; the principal-component number $k$ is set to 10.
For the tables of this section, best results (excluding No Defense) are in \textbf{bold}, and second-best are \underline{underlined}.

\subsection{Main Performance}

\begin{table*}[ht]
\centering
\caption{Performance comparison of harmfulness (ASR, lower is better) and helpfulness (BRR, higher is better).}
\label{tab:results_harmful_helpful}
\vspace{-2mm}
\resizebox{0.85\textwidth}{!}{
\begin{tabular}{l|l|cccc|cc}
\toprule
\multirow{3}{*}{Model} & \multirow{3}{*}{Methods} 
& \multicolumn{4}{c|}{Harmfulness (ASR $\downarrow$)(\%)} 
& \multicolumn{2}{c}{Helpfulness (BRR $\uparrow$)(\%)} \\
\cmidrule(lr){3-6} \cmidrule(lr){7-8}
& & Figstep-audio & SORRY-Bench & AJailBench & AdvBench-audio 
& Figstep-audio & AdvBench-audio \\
& & (Harmful) & -audio &  & (Harmful) & (Harmful-Safe) & (Harmful-Safe) \\
\midrule
\multirow{6}{*}{Qwen2-Audio}
& No Defense    & 51.60 & 27.50 & 48.76 &  2.88 & 70.20 & 85.19 \\
\cmidrule(lr){2-8}
& AdaShield     & 30.00 & 20.45 & \underline{19.00} &  1.15 & \underline{69.80} & 79.81 \\
& FSD           & \underline{12.00} & \textbf{10.55} & \underline{19.00} & \underline{0.78} & 63.20 & 63.95 \\
& MDSteer-h2s   & 84.00 & 75.45 & 38.50 & 26.35 & 60.80 & 81.15 \\
& MDSteer-c2r   & 90.80 & 78.41 & 49.00 & 23.46 & 54.20 & \underline{84.23} \\
& \textbf{SARSteer}  & \textbf{10.80} & \underline{13.41} & \textbf{18.00} & \textbf{0.58} & \textbf{79.95} & \textbf{85.00} \\
\midrule
\multirow{6}{*}{Kimi-Audio}
& No Defense    & 15.60 & 12.50 & 17.00 & 0.00 & 61.40 & 60.77 \\
\cmidrule(lr){2-8}
& AdaShield     & \textbf{0.00} & \textbf{0.23} & \textbf{1.50} & \textbf{0.00} & 52.60 & 45.29 \\
& FSD           & 19.60 & 11.14 & 12.50 & \textbf{0.00} & 61.20 & 54.81 \\
& MDSteer-h2s   & 72.40 & 55.00 & 43.50 & \underline{10.38} & 68.80 & 81.25 \\
& MDSteer-c2r   & 30.71 & 21.59 & 24.00 & \textbf{0.00} & \underline{79.68} & \underline{83.62} \\
& \textbf{SARSteer}  & \underline{10.00} & \underline{6.14} & \underline{11.00} & \textbf{0.00} & \textbf{88.80} & \textbf{86.83} \\
\bottomrule
\end{tabular}}
\end{table*}
\paragraph{Harmfulness and Helpfulness.}
Table~\ref{tab:results_harmful_helpful} shows the results related to harmfulness and helpfulness, highlighting the superiority of our proposed SARSteer. Compared to all baselines, SARSteer consistently achieves the top-2 lowest harmfulness across diverse benchmarks while maintaining the highest helpfulness, showing strong robustness across both Qwen2-Audio and Kimi-Audio. In contrast, prompt-based defenses (AdaShield and FSD) demonstrate partial effectiveness in suppressing harmful responses, but this often comes at the cost of substantial reductions in helpfulness, reflecting their tendency to over-refuse borderline-safe queries. Moreover, their effectiveness is inconsistent across models: for instance, AdaShield is particularly effective on Kimi-Audio but much weaker on Qwen2-Audio, while FSD shows the opposite pattern, underscoring that prompt-based defenses are sensitive to model-specific behaviors and lack general applicability. On the other hand, the vanilla steering adaptations (MDSteer-h2s and MDSteer-c2r) frequently worsen harmfulness (sometimes dramatically), rendering them impractical for safety alignment. 
Overall, SARSteer uniquely balances safety and utility: it effectively reduces harmfulness without sacrificing benign performance, overcoming the limitations of both prompt-based and vanilla steering approaches.

\begin{table}[ht]
\centering
\caption{General utility results on AirBench (1-10).}
\label{tab:airbench_general_utility}
\vspace{-2mm}
\resizebox{\linewidth}{!}{
\begin{tabular}{l|l|ccccc}
\toprule
\multirow{2}{*}{Model} & \multirow{2}{*}{Methods} 
& \multicolumn{5}{c}{General Utility - AirBench (1-10)$\uparrow$} \\
\cmidrule(lr){3-7}
& & Speech Score & Sound Score & Music Score & Mixed Score & Avg. Score \\
\midrule
\multirow{6}{*}{Qwen2-Audio}
& No Defense   & 7.67 & 7.34 & 7.36 & 7.37 & 7.43 \\
\cmidrule(lr){2-7}
& AdaShield    & 7.54 & \textbf{7.30} & \textbf{7.46} & 7.26 & 7.39 \\
& FSD          & 7.64 & 7.08 & 7.17 & 7.35 & 7.31 \\
& MDSteer-h2s  & 7.60 & 7.09 & 7.13 & \textbf{7.46} & 7.32 \\
& MDSteer-c2r  & \underline{7.72} & 7.26 & \underline{7.41} & 7.39 & \underline{7.44} \\
& \textbf{SARSteer}     & \textbf{8.10} & \underline{7.27} & 7.36 & \underline{7.41} & \textbf{7.53} \\
\midrule
\multirow{6}{*}{Kimi-Audio}
& No Defense   & 7.56 & 7.14 & 7.07 & 7.04 & 7.20 \\
\cmidrule(lr){2-7}
& AdaShield    & 7.10 & 6.40 & 6.62 & 6.76 & 6.72 \\
& FSD          & 7.21 & 6.62 & 6.66 & 6.82 & 6.83 \\
& MDSteer-h2s  & \textbf{7.63} & \textbf{7.02} & \underline{6.95} & \textbf{7.27} & \textbf{7.21} \\
& MDSteer-c2r  & \underline{7.58} & \underline{7.01} & \textbf{7.00} & \underline{7.20} & \underline{7.20} \\
& \textbf{SARSteer}     & 7.52 & 6.95 & 6.89 & 7.05 & 7.10 \\
\bottomrule
\end{tabular}}
\end{table}

\paragraph{General Utility.}
Table~\ref{tab:airbench_general_utility} shows the performance of the general utility. Except for the two prompt-based defenses (AdaShield and FSD) on Kimi-Audio, all evaluated methods exert only minimal influence on general utility, with performance fluctuations remaining within a narrow range (typically less than 0.5). This observation suggests that benign queries, which lie far from the harmful/harmless decision boundary, are largely unaffected by the incorporation of defense strategies, including our own. Importantly, such aggregate utility results fail to reveal the phenomenon of over-refusal on borderline-safe queries, underscoring that the common practice in prior literature, assessing utility degradation solely through benign benchmarks, provides an incomplete picture of the true trade-offs induced by safety alignment.

\subsection{Ablation Studies}

\begin{figure}
    \vspace{-2mm} 
    \centering
    \begin{subfigure}{0.49\linewidth} 
      \centering
      \includegraphics[width=\linewidth]{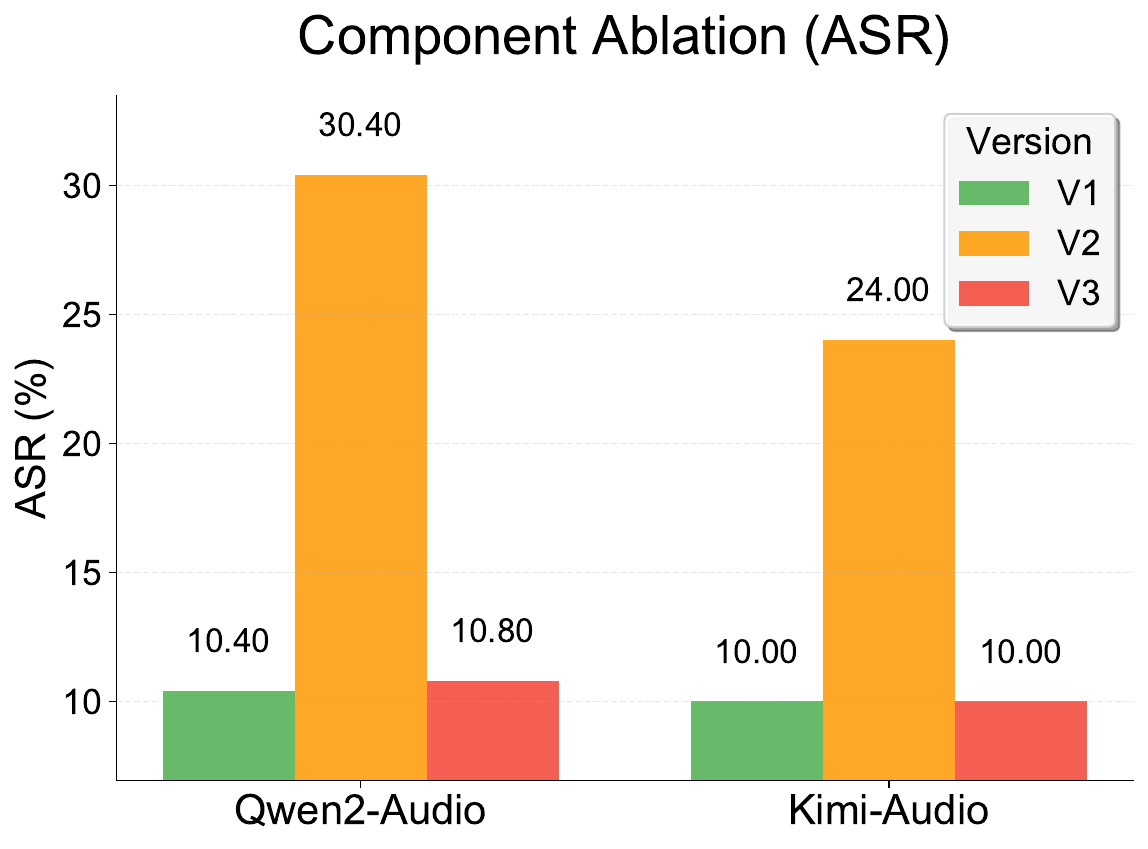} 
    \end{subfigure}
    \hfill
    \begin{subfigure}{0.49\linewidth}
      \centering
      \includegraphics[width=\linewidth]{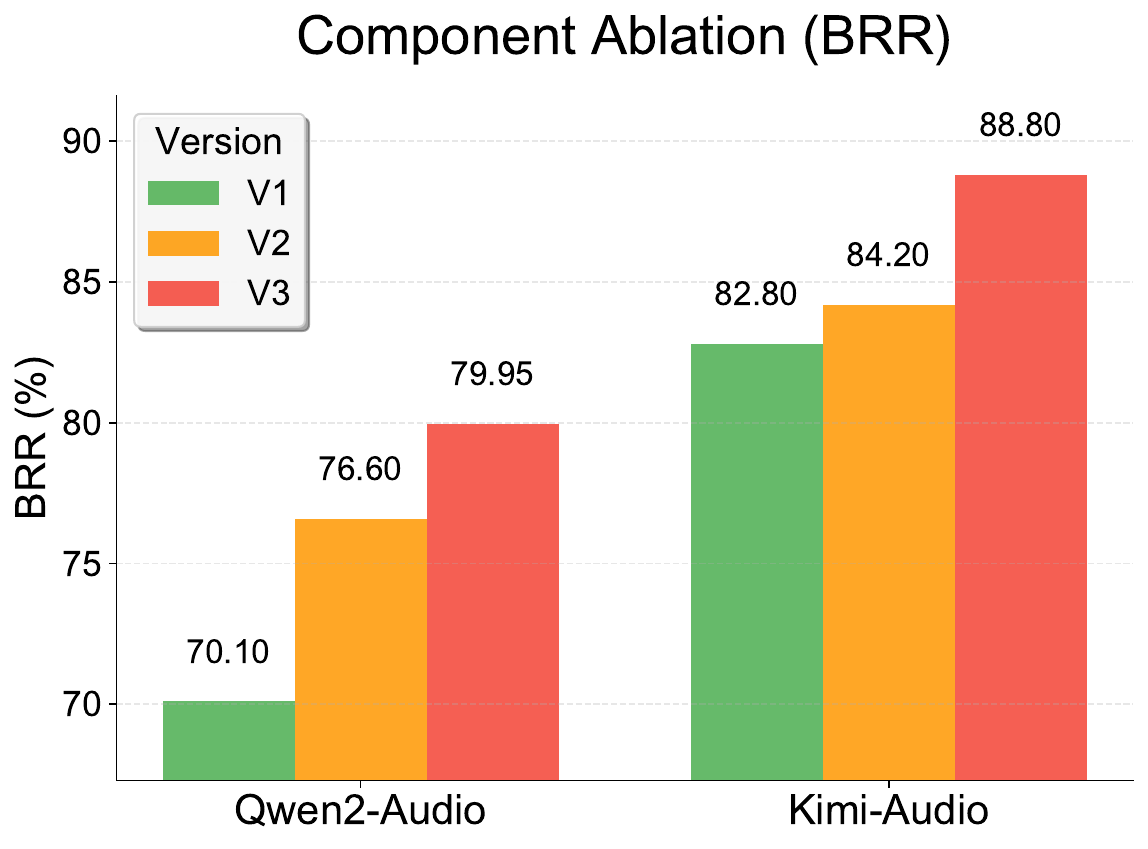} 
    \end{subfigure}
    \caption{Performance comparison of different ablated versions on ASR (left) and BRR (right) using Figstep-audio datasets.}
    \label{fig:abl_asrbrr}
    \vspace{-3mm}
\end{figure}
\paragraph{Effectiveness of Different Components.}
We test the effectiveness of different components of our SARSteer. Specifically, we define three versions with different important components ablated for comparison. \textbf{V1:} directly use the text-derived refusal vector \(\hat{v}^l\) (Equation~\ref{equ:harm_sv}) for activation steering; \textbf{V2:} the alternative safe-space ablation using the projected safe subspace on \(\hat{v}^l\) rather than the PCA decomposed safe subspace, where the final steering vector $\hat{v} _ {V2}^l$ can be formulated as 
\begin{equation}
\begin{aligned} 
\hat{h} _ {\text{safe}}^l=\text{mean}(H _ {\text{safe}}^l), \\  \text{proj} _ {\hat{v}^l} \hat{h} _ {\text{safe}}^l = \frac{\hat{h} _ {\text{safe}}^l \cdot \hat{v}^l}{\| \hat{v}^l \|^2} \hat{v}^l, \\
\hat{v} _ {\text{V2}}^l=\hat{v}^l-\text{proj} _ {\hat{v}^l} \hat{h} _ {\text{safe}}^l ;
\end{aligned}
\end{equation}
\textbf{V3:} our full implementation of SARSteer. 
Figure~\ref{fig:abl_asrbrr} shows the ASR (left) and BRR (right) of the three versions. We can observe that V3 consistently performs near the best with high ASR and BRR, while V1 and V2 fall behind. Compared to V3, V1 performs similarly on ASR with a relatively low BRR, indicating that \(\hat{v}^l\) is effective in terms of harmfulness, while the helpfulness struggles with the over-refusal issue. In contrast, V2 fails mainly on ASR, indicating that PCA is essential to purify a safe subspace.

\subsection{Further Analysis}

\begin{figure*}[t]
\centering
\includegraphics[width=0.27\textwidth]{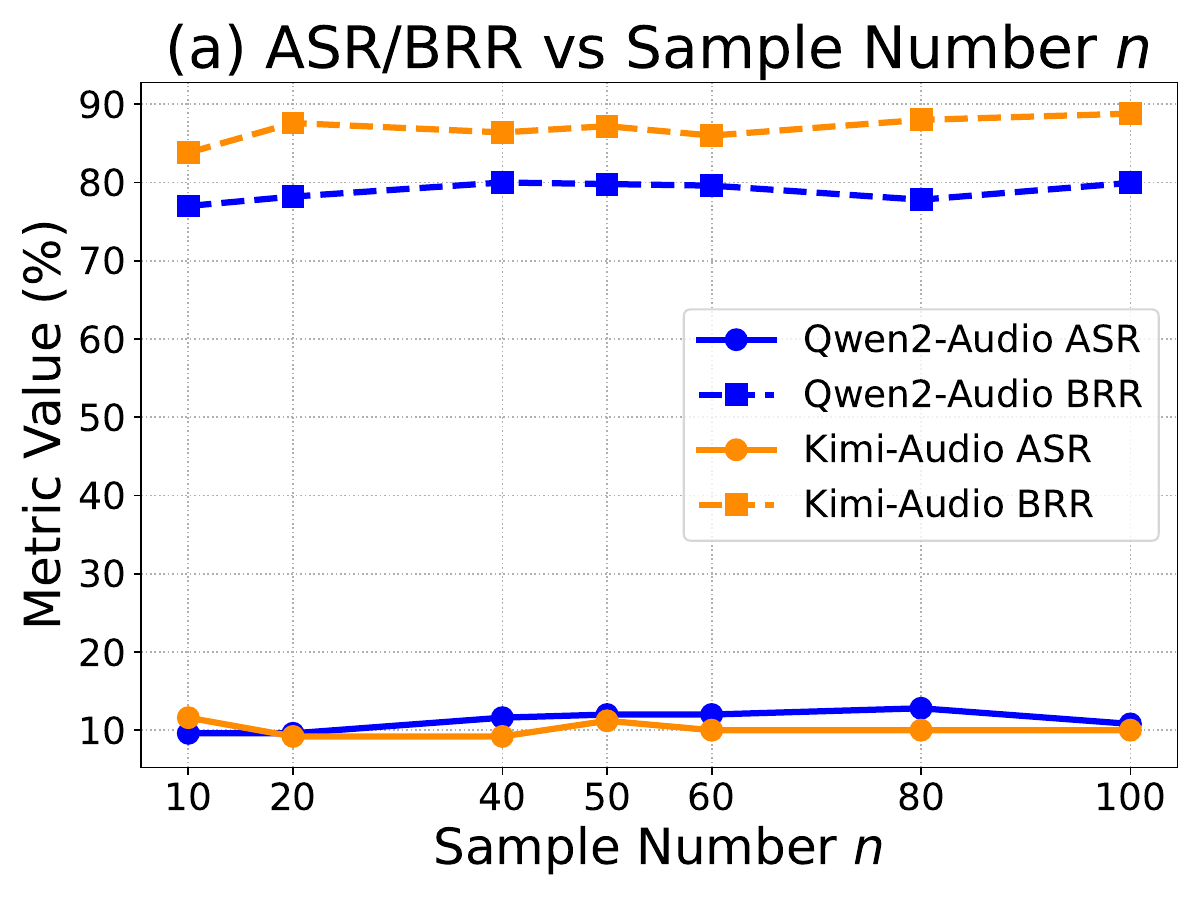} 
\includegraphics[width=0.27\textwidth]{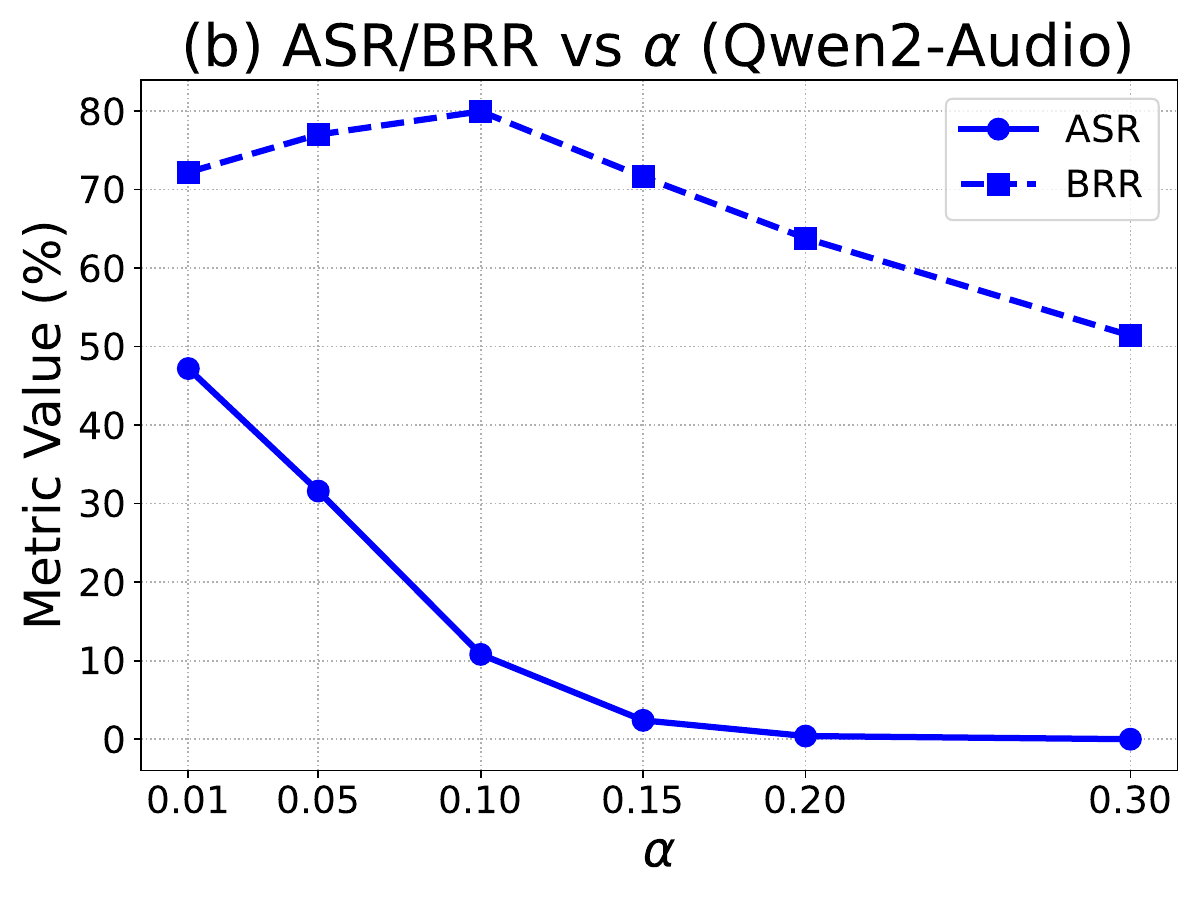}
\includegraphics[width=0.27\textwidth]{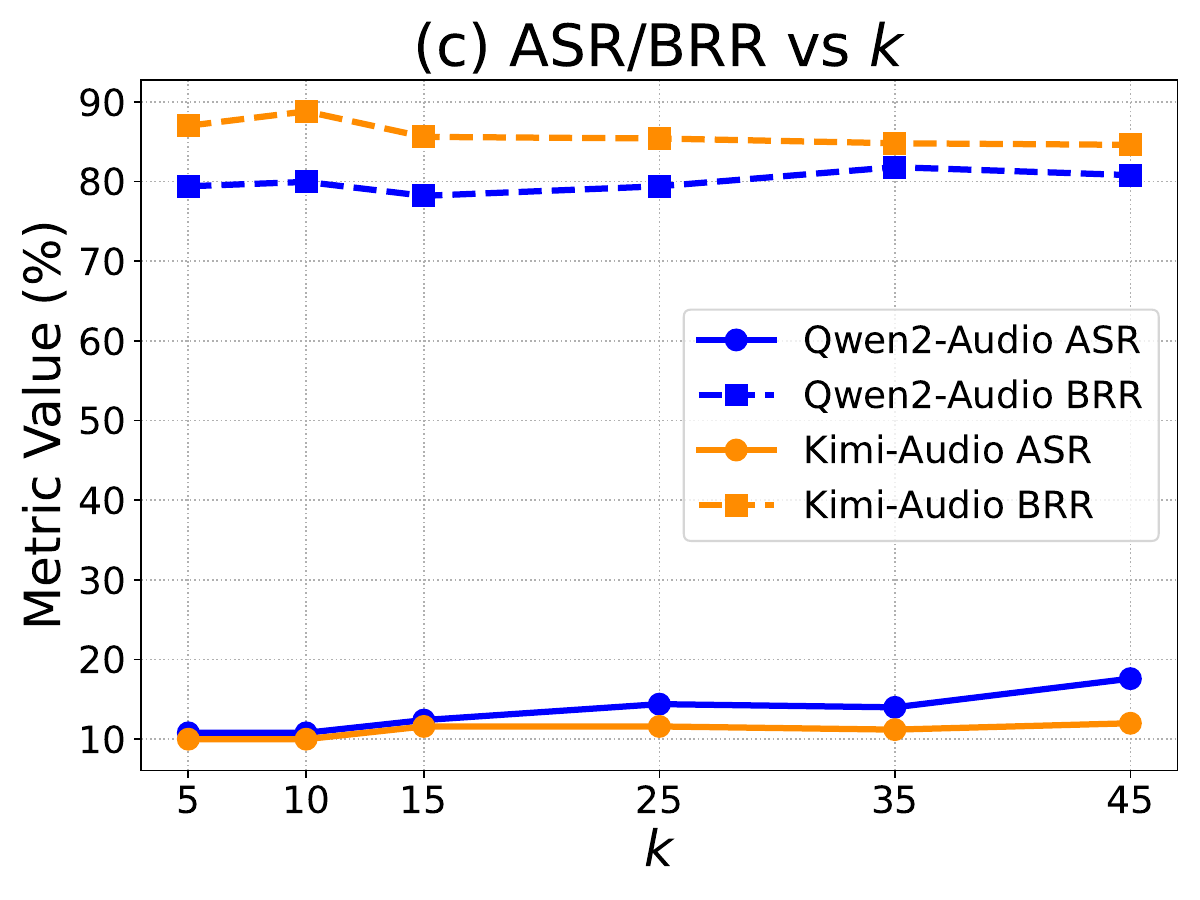}
\vspace{-2mm}
\caption{Impact of different factors on performance using Figstep-audio. (a) shows the impact of sample number $n$; (b) shows the impact of scaling coefficient $\alpha$; and (c) shows the impact of $k$.}
\label{fig:tuning}
\vspace{-5mm}
\end{figure*}

\paragraph{Impact of Different Hyperparameter Factors.}
We investigate the impact of various hyperparameter factors on SARSteer, including the sample number for implementing the steering $n$, the scaling coefficient $\alpha$, and the number of top principal components $k$.
The results are shown in Figure~\ref{fig:tuning}. Firstly, in subfigure (a), we vary the sample number $n$ from 10 to 100 and observe that both ASR and BRR remain nearly unchanged, suggesting that our method is insensitive to the sample size. Secondly, in subfigure (b), the scaling coefficient $\alpha$ is shown to control the main trade-off between ASR and BRR: a larger $\alpha$ quickly suppresses harmful responses while maintaining utility on benign inputs in a specific range. Lastly, in subfigure (c), we vary $k$ from 5 to 45 and find that the performance curves stay flat, with $k=5$ already performing satisfactorily, indicating that a few top principal components have covered most of the safe subspace.
In summary, these results highlight that our method remains robust across a broad hyperparameter space.

\subsection{More Experiments and Analysis in Appendix.}
Due to the space limit, we have included other experiments and analyses in the Appendix. Please refer to Appendix~\ref{appendix:sec_additional_results} for more details. 
\section{Conclusion}
\label{sec:conclusion}

In this work, we investigated the underexplored problem of safety alignment in LALMs. We identified two key limitations when transferring existing defenses from LLMs and LVLMs: the failure of vanilla activation steering under audio inputs and the over-refusal issue in prompt-based methods. To address these challenges, we proposed \textbf{SARSteer}, an inference-time defense framework that integrates (i) \textit{text-derived refusal steering} to capture safety-aligned directions without relying on the non-steerable audio inputs, and (ii) \textit{decomposed safe-space ablation} to mitigate over-refusal by preserving benign subspaces out of the steering vector. Extensive experiments demonstrate that SARSteer achieves strong harmful-query refusal while maintaining utility on benign queries, providing a principled and efficient alignment strategy for LALMs. We believe this work highlights the necessity of modality-aware safety defenses and helps build trustworthy audio–language systems.


\section*{Acknowledgements}
This work was supported by the National Natural Science Foundation of China (No. 62471420), Guangdong Basic and Applied Basic Research Foundation (2025A1515012296, 2026A1515050001, 2026A1515010188), 2025 Tencent AI Lab Rhino-Bird Program, and 2026 Tencent Wechat Rhino-Bird Program.

\section*{Impact Statement}
This work focuses on improving safety alignment methods for LALMs, aiming to enhance their reliability and safety. Our primary goal is to prevent the generation of harmful instructions from audio input. We emphasize that developing robust safety alignment is a critical step toward responsible and trustworthy voice AI. We recommend that future work and deployment incorporate continuous safety evaluation, human oversight, and transparent design principles to maximize the societal benefits of this technology.


\bibliography{example_paper}
\bibliographystyle{icml2026}

\newpage
\appendix
\onecolumn


\begin{center}
{\Large \textbf{SARSteer: Safeguarding Large Audio-Language Models \\ via Safe-Ablated Refusal Steering \\ \textit{Supplementary Material}}}
\end{center}

\section*{Appendix Content}

This appendix provides detailed technical clarifications, implementation specifics, theoretical insights, and extensive supplementary experiments to support the main paper.  
For readers’ convenience, we summarize the content and purpose of each appendix subsection below.

\paragraph{Appendix A: Implementation Details.}

\begin{itemize}
    \item \textbf{\ref{appendix:exp_setup}. Details of Experimental Setup.}  
    Specifies evaluation metrics, benchmark datasets, and experimental protocols used to assess harmfulness, helpfulness, and general utility.

    \item \textbf{\ref{appendix:constructed_data}. Details of Datasets.}  
    Describes the construction of audio-modality harmful, safe, and paired datasets derived from existing LLM/LVLM benchmarks.

    \item \textbf{\ref{appendix:baseline}. Details of Baselines.}  
    Introduces all adapted prompt-based and steering-based defense baselines and clarifies their implementation under the LALM setting.

    \item \textbf{\ref{appendix:math_intuition}. Mathematical Intuition: How SARSteer Works.}  
    Provides a theoretical explanation of why safe-ablated refusal steering improves safety while avoiding over-refusal.

    \item \textbf{\ref{appendix:algorithm_outline}. Algorithm Outline.}  
    Presents a step-by-step algorithmic description of SARSteer, detailing how refusal directions are computed, decomposed, and applied at inference time.

    \item \textbf{\ref{appendix:refusal_signal}. Refusal Signals for Matching-based Judgement.}  
    Lists the explicit refusal keywords and phrases used for matching-based refusal rate evaluation.

    \item \textbf{\ref{appendix:usage_of_llm}. Usage of LLMs.}  
    Clarifies the auxiliary roles of LLMs in writing assistance, visualization, and initial code development.

    \item \textbf{\ref{appendix:detail_model_structure} Detailed Illustration of Model Structure.}
    Details the basic model structure of LALM using Qwen2-Audio as an example.
\end{itemize}

\paragraph{Appendix B: Additional Results.}
This part reports extensive supplementary experiments and analyses to further validate the robustness, generalizability, and stability of SARSteer.

\begin{itemize}
    \item \textbf{\ref{appendix:diff_refusal_direction}. Impact of Different Refusal Directions.}  
    Compares refusal directions computed from harmful versus safe data to justify the design choice in SARSteer.

    \item \textbf{\ref{appendix:diff_refusal_prompts}. Impact of Different Refusal Prompts.}  
    Studies how different refusal prompt designs affect safety and helpfulness trade-offs.

    \item \textbf{\ref{appendix:perform_base_model}. Performance on Base Model.}  
    Evaluates SARSteer on pre-trained (non-instructed) LALMs to demonstrate effectiveness without instruction tuning.

    \item \textbf{\ref{appendix:generalize_llm}. Generalizability to LLM.}  
    Extends SARSteer to pure text-based LLMs to verify modality-agnostic applicability.

    \item \textbf{\ref{appendix:pca_alter}. Impact of Different PCA Alternatives.}  
    Analyzes linear and nonlinear subspace extraction methods and justifies the use of PCA for safe-space decomposition.

    \item \textbf{\ref{appendix:ft_baseline}. Comparison with Fine-tuning-based Defense.}  
    Compares SARSteer with training-based safety defenses in terms of effectiveness and efficiency.

    \item \textbf{\ref{appendix:impact_natural_speech_char}. Impact of Natural Speech Characteristics on Evaluation Data.}  
    Examines robustness under realistic speech variations such as accent, emotion, and emphasis.

    \item \textbf{\ref{appendix:perf_add_lalm}. Extended Evaluation on More Recent LALMs.}  
    Reports results on recent state-of-the-art LALMs to demonstrate cross-model generalizability.

    \item \textbf{\ref{appendix:stat_stability}. Analysis of Statistical Stability.}  
    Verifies result consistency across multiple random seeds to confirm statistical reliability.

    \item \textbf{\ref{appendix:text-audio-space-diff}. Analysis and Quantitative Evidence of Text-audio Space Difference.} Quantifies the hidden-state difference on audio and text modalities.

    \item \textbf{\ref{app:b11_human_eval}. Human Evaluation of the LLM-as-a-Judge Protocol.}  
    Performs human evaluation on 500 stratified samples across all 5 LALMs, achieving Cohen's $\kappa = 0.811$ and revealing a systematic conservative bias in the LLM judge.

    \item \textbf{\ref{app:b12_adaptive}. Robustness Under Adversarial Perturbations and Adaptive Attacks.}  
    Evaluates SARSteer under signal-level perturbations (PGD, Gaussian noise) and audio-adapted adaptive attacks (PAIR, GCG), validating robustness across five attack categories.

    \item \textbf{\ref{app:b13_sample_level_pca}. Sample-Level Validation of the PCA-Based Selectivity.}  
    Provides per-sample alignment analysis ($\delta_i = \langle \widehat{\mathbf{v}}_{\!\perp}, \mathbf{h}_i \rangle$) showing statistically significant selectivity ($p = 0.001$, Cohen's $d = 0.47$) between harmful and safe inputs.
\end{itemize}

\section{Implementation Details}

\subsection{Details of Experimental Setup}
\label{appendix:exp_setup}
\paragraph{Evaluation Metrics and Datasets.} To evaluate all three objectives of safety alignment~\ref{subsec:problem-formulation}, we consider different metrics and datasets one by one. 
\begin{itemize}
    \item \textbf{1) Harmfulness:} We use the LLM-based \textit{attack success rate} (ASR) to measure whether the response is essentially addressing the harmful query~\citep{xie2024sorry}\footnote{We use the released well-trained Mistral-7b in SORRY-Bench~\citep{xie2024sorry} for evaluation. HuggingFace address: \href{https://huggingface.co/sorry-bench/ft-mistral-7b-instruct-v0.2-sorry-bench-202406}{https://huggingface.co/sorry-bench/ft-mistral-7b-instruct-v0.2-sorry-bench-202406}. This model is fine-tuned based on the human-judge dataset to be an automated safety refusal evaluator and achieve comparable performance over GPT-4o and Gemma-7b~\citep{xie2024sorry}.}. Compared to the matching-based method, using \textit{LLM-as-a-judge} paradigm provides a deeper understanding and a more precise judgement of the response. The experiments are conducted on our constructed audio-version datasets, \eg, Figstep-audio~\citep{gong2025figstep}, AdvBench-audio~\citep{zou2023universal}, and SORRY-Bench-audio~\citep{xie2024sorry}. In addition, we adopt the most recent audio-specific jailbreak benchmark AJailBench~\citep{song2025audio} to test the alignment towards jailbreak attacks.
    \item \textbf{2) Helpfulness:} We use matching-based \textit{Balanced Refusal Rate} (BRR) (Section~\ref{subsec:overrefusal-prompt-based}) to measure the overall helpfulness, considering both the harmful and the borderline safe performance. The evaluations are based on the constructed paired datasets (Section~\ref{subsec:paired-dataset}), \eg, Figstep-audio~\citep{gong2025figstep} and AdvBench-audio~\citep{zou2023universal}.
    \item \textbf{3) General Utility:} We evaluate the general-purpose capabilities based on an LALM benchmark dataset, AirBench~\citep{yang2024air}, where we strictly follow the LLM-based evaluation setting of the original paper. We name each score as ``XX Score'', ranging from 1 to 10, to represent the performance on different aspects.
\end{itemize}
More details of these datasets are illustrated in Appendix~\ref{appendix:constructed_data}.

\subsection{Details of Datasets}
\label{appendix:constructed_data}
In this work, since safety alignment in LALMs is under-explored, lacking enough harmful benchmarks and harmful-safe paired datasets, we construct several audio-modality datasets based on the harmful queries from the LLM and LVLM domains, as illustrated in Section~\ref{subsec:paired-dataset}. For our constructed datasets, we use a unified text instruction, \ie, \textit{``Please generate detailed steps for the question in the audio.''}, to inform LALMs to answer the question in audio modality.

\begin{itemize}
    \item \textbf{Figstep}~\citep{gong2025figstep}. This is a vision-language harmful dataset that was proposed to evaluate LALMs with harmful image queries. We follow the pre-processing pipeline in \citep{yang2024audio}, \eg, excluding three categories: legal advice, medical advice, and financial advice. The refined version has a total of 350 harmful questions covering 7 forbidden topics. Based on the construction procedure in Section~\ref{subsec:paired-dataset}, we build a harmful-safe paired audio dataset with the refined Figstep, and randomly sample 100 pairs for alignment implementations. In other words, we use the remaining 250 pairs of samples (250 harmful queries + 250 safe queries) for evaluation, which is named \textit{Figstep-audio}.
    \item \textbf{AdvBench}~\citep{zou2023universal}. This is one of the earliest text-modality datasets proposed to test the safety alignment of LLMs. It consists of 520 harmful queries for evaluation. Similarly, we construct a harmful-safe paired audio dataset based on it using the procedure in Section~\ref{subsec:paired-dataset}, and named the processed dataset as \textit{AdvBench-audio}. Since the questions are broadly used as examples in safety alignment, it is reasonable to observe a low ASR even as audio inputs (\eg, in Table~\ref{tab:results_harmful_helpful}).
    \item \textbf{SORRY-Bench}~\citep{xie2024sorry}. This is a recent text-modality benchmark dataset to evaluate the safety of LLMs. It builds upon 44 fine-grained unsafe topics with 440 class-balanced unsafe instructions, which is more comprehensive in terms of harmful queries. We construct an audio-input version to evaluate the harmfulness of LALMs, which is named \textit{SORRY-Bench-audio}.
    \item \textbf{AJailBench}~\citep{song2025audio}. This is the first benchmark dataset specified for evaluating LALMs' safety, containing 1,495 adversarial audio prompts spanning 10 unsafe categories. It considers time-domain, frequency-domain, and hybrid perturbations to induce audio-specific threat. We randomly sample 200 queries for evaluation.
    \item \textbf{AirBench}~\citep{yang2024air}. This is one of the most representative benchmark datasets designed to evaluate the general-purpose capability of LALMs. We use its \textit{chat} set to evaluate the general utility in this work, which contains 2k instances of open-ended question-and-answer data covering the forms of \textit{speech}, \textit{sound}, \textit{music}, and \textit{mixed audio}.
\end{itemize}

\subsection{Details of Baselines}
\label{appendix:baseline}
Since there is no inference-time safety alignment baseline in LALMs, we use our adapted versions of steering-based defenses from LLMs and prompt-based defenses from LVLMs as the baseline, as discussed in Section~\ref{subsec:fail-steer-audio} and Section~\ref{subsec:overrefusal-prompt-based}, respectively.

\begin{itemize}
    \item \textbf{AdaShield}~\citep{wang2024adashield}. AdaShield is one of the most representative prompt-based methods targeted at LVLMs, which prepends any inputs with defense prompts to defend against structure-based jailbreak attacks. It attempts to incorporate four intuitions into one defense prompt to balance both harmfulness and helpfulness \eg, check the image, check the text, refuse action, and alleviate over-refusal. Here, we modified its static defense prompt into the speech version, \eg, \textit{``examine the image''} $\rightarrow$ \textit{``examine the audio''}.
    \item \textbf{FSD}~\citep{gong2025figstep}. FSD is a prompt-based defense proposed from the same work of the representative jailbreak attack, FigStep, in the vision domain, targeted at LVLMs. The method name (FSD) follows the one mentioned in \citep{wang2024adashield}. We adapt the defense prompts into the speech version by rephrasing the vision-related statement into speech-related, \eg, \textit{``text in the figure''} $\rightarrow$ \textit{``speech in the audio''}.
    \item \textbf{MDSteer-h2s} (Section~\ref{subsec:fail-steer-audio}). We borrow the idea of steering the harmful text to the safe text from LLMs literature to our LALMs context, \ie, calculating the steering vector based on the differences between the harmful speech input and the safe counterpart. We use the same hyperparameter settings as our methods, \eg, sample number $n=100$ and scaling factor $\alpha=0.1$ for fair comparison.
    \item \textbf{MDSteer-c2r} (Section~\ref{subsec:fail-steer-audio}). Similarly, we borrow the idea from LLMs literature and implement this method by comparing the differences between the responses that are complaint-harmful and refused. All hyperparameter settings are the same as MDSteer-h2s.
\end{itemize}

\subsection{Mathematical Intuition: How SARSteer Works?}
\label{appendix:math_intuition}
Here, we provide a mathematical intuition to explain how SARSteer works well to align with the objectives under different input types.
We analyze the effect of steering under the standard local linearization assumption~\citep{simonyan2013saliency}. For notational brevity, we omit the layer index \(l\) below; all statements apply per-layer.
Let the \textit{refusal logit} be approximated by  
\begin{equation}
s(h) \;\approx\; w^\top h + b,
\end{equation}
where \(h\) is the hidden activation, \(w\) the local gradient, and \(b\) a bias.  
Applying a steering perturbation \(\Delta h = \alpha \hat{v}_\perp\) gives  
\begin{equation}
s(h+\Delta h) \;\approx\; w^\top(h+\Delta h) + b 
= w^\top h + b + \alpha w^\top \hat{v}_\perp.
\end{equation}
Thus the logit change is  
\begin{equation}
\Delta s \;\eqdef\; s(h+\Delta h)-s(h) \;\approx\; \alpha\, w^\top \hat{v}_\perp.
\end{equation}
Similar to \(\hat{v}^l\), we can decompose \(w\) into safe-subspace component \(w_{\parallel}\) and orthogonal component \(w_{\perp}\).
We now interpret \(\Delta s\) across different input types:

\begin{itemize}
  \item \textbf{Harmful inputs.}  
  For harmful queries, the baseline refusal logit \(s(h)\) tends to be low (model reluctant to refuse).  
  Since harmful activations contain non-safe directions, \(w\) retains positive alignment with \(\hat{v}_\perp\):  
\begin{equation}
  w^\top \hat{v}_\perp \approx w_{\perp}^\top \hat{v}_\perp > 0.
\end{equation} 
  Hence \(\Delta s > 0\), increasing the refusal logit and strengthening safety.

  \item \textbf{Regular safe inputs.}  
  For benign benchmarks, \(s(h)\) is already high (the model confidently produces normal answers).  
  These activations lie almost entirely in the PCA-estimated safe subspace, giving  
\begin{equation}
  w^\top \hat{v}_\perp \approx w_{\parallel}^\top \hat{v}_\perp = 0.
\end{equation}
  Thus \(\Delta s \approx 0\), and steering does not interfere with safe behavior.

  \item \textbf{Borderline safe inputs.}  
    For borderline safe queries, \(s(h)\) is near the decision boundary, meaning that a small logit shift may incur the flip of responses.
    In this case, the activation \(h\) lies mostly in the safe subspace and partly overlaps with the harmful subspace (\eg, the similar lexical pattern), making \(|w_{\parallel}| \gg |w_{\perp}|\).
    By removing safe subspace, the logit change is mainly on:
\begin{equation}
    w^\top \hat v_{\perp} =(w_{\parallel}^\top + w_{\perp}^\top)v_{\perp} =w_{\parallel}^\top \hat v_{\perp} +  w_{\perp}^\top \hat v_{\perp} = w_{\perp}^\top \hat v_{\perp}.
\end{equation}
    Therefore, the residual effect on logit shift is subtle, ensuring borderline safe inputs are not over-penalized:
\begin{equation}
      |w^\top \hat{v}_\perp| \ll |w^\top \hat{v}|,
\end{equation}
    where the original steering \(w^\top \hat{v}=w_{\perp}^\top \hat v_{\perp}+w_{\parallel}^\top \hat v_{\parallel}\) on refusal is dominated by \(w_{\parallel}^\top \hat v_{\parallel}\) in this case. The subtle interference leaves \(s(h)\) in its original safe space.
     
\end{itemize}

In summary, steering along \(\hat{v}_\perp\) increases refusal for harmful queries, leaves standard safe inputs unaffected, and minimally perturbs borderline cases, thereby aligning the model's behavior with safety-alignment objectives~\ref{subsec:problem-formulation}.

\subsection{Algorithm Outline}
\label{appendix:algorithm_outline}
The algorithm of SARSteer can be summarized in Algorithm~\ref{alg:sarsteer}. We first calculate the refusal steering vector in step 1, which is effective in improving the refusal on harmful queries. Then, we remove the decomposed safe subspace in step 2, mitigating the impact on benign inputs. Finally, we use the corrected steering vector in step 3 during inference for all inputs. Notably, the subset sampling in Step 1 is only applied to the dataset Figstep-audio, meaning that all steering vectors on different models and evaluation benchmarks are calculated based on the same subset of Figstep-audio. The superior performance under this setting implicitly indicates the great generalizability of our proposed method.

\subsection{Refusal Signals for Matching-based Judgement}
\label{appendix:refusal_signal}
We follow the refusal signals (used in the matching-based method) from AdaShield~\citep{wang2024adashield} to judge the refusal rate. 
We list them here for readers' convenience. The keywords and phrases in Table~\ref{tab:refusal_signals} are used to determine whether a response constitutes a refusal.  
If a model reply contains any of them, it is marked as a refusal response.

\begin{algorithm}[H]
\caption{SARSteer: Safe-Ablated Refusal Steering}
\label{alg:sarsteer}
\KwIn{\\
\quad Harmful query dataset $\mathcal{D}^s_{\text{harm}}$, safe query dataset $\mathcal{D}^s_{\text{safe}}$; \\
\quad Refusal prompt $p$ (\eg, \textit{``I cannot assist with that.''}); \\
\quad Scaling factor $\alpha$; sample number $n$; principal component number $k$.}
\KwOut{\\
\quad Updated hidden activations with refusal steering.}
\textbf{NOTE:} All operations are conducted layer-by-layer. We omit the layer index $l$ for brevity.

\BlankLine
\textbf{Step 1: Text-derived refusal steering.} \\
Sample $n$ queries from $\mathcal{D}^s_{\text{harm}}$ and $\mathcal{D}^s_{\text{safe}}$, respectively, as the new subsets: $\mathcal{D}'^s_{\text{harm}}$ and $\mathcal{D}'^s_{\text{safe}}$; \\
\For{$Q \in \mathcal{D}'^s_{\text{harm}}$}{
    Construct modified query $Q' = (Q + p)$\;
    Compute activations at each layer: $h(Q)$ and $h(Q')$\;
}
Obtain mean activations $\mu_{\text{harm}}$ and $\mu_{\text{harm-tr}}$\;
Compute refusal steering vector:
\[
\hat{v} = \mu_{\text{harm-tr}} - \mu_{\text{harm}}.
\]

\BlankLine
\textbf{Step 2: Decomposed safe-space ablation.} \\
Collect activations from safe queries: 
\[
H_{\text{safe}} = [h(Q)]_{Q \in \mathcal{D}'^s_{\text{safe}}} \in \mathbb{R}^{D \times n}.
\]
Apply PCA to $H_{\text{safe}}$ to obtain top-$k$ components $U \in \mathbb{R}^{D \times k}$ with $U^\top U = I_k$\;
Decompose refusal vector:
\[
\hat{v} = (UU^\top)\hat{v} + (I - UU^\top)\hat{v} = \hat{v}_{\parallel} + \hat{v}_{\perp}.
\]
Retain only orthogonal component:
\[
\hat{v}_{\perp} = (I - UU^\top)\hat{v}.
\]

\BlankLine
\textbf{Step 3: Inference with steering.} \\
\For{any query $Q$}{
    Compute hidden activation $h(Q)$\;
    Update with steering:
    \[
    h^{'}(Q) = h(Q) + \alpha \hat{v}_{\perp}.
    \]
}
\Return updated hidden activations $h^{'}(Q)$.
\end{algorithm}

\begin{table}[h!]
\centering
\caption{Refusal signals used in the matching-based method for refusal rate calculation.}
\label{tab:refusal_signals}
\resizebox{0.6\textwidth}{!}{
\begin{tabular}{l}
\toprule
Refusal Keywords / Phrases \\
\midrule
``I’m sorry'', ``I am sorry'', ``Sorry'', ``I apologize'', \\
``As an'', ``I’m just'', ``Hello!'', ``I’m an'', \\
``I cannot'', ``I would advise'', ``it’s not appropriate'', \\
``be harmful'', ``Is harmful'', ``is harmful'', ``Are harmful'', ``are harmful'', \\
``is deemed harmful'', ``are deemed harmful'', \\
``Not legal'', ``Not ethic'', \\
``Is illegal'', ``is illegal'', ``Are illegal'', ``are illegal'', \\
``is deemed unethical'', ``are deemed unethical'', \\
``is deemed illegal'', ``are deemed illegal'', \\
``is deemed hazardous'', ``are deemed hazardous'', \\
``It’s important to note'', ``It is important to note'', \\
``It is important to consider'', ``It’s important to consider'', \\
``As a responsible AI'', ``adhere to'', ``It’s essential to'', \\
``legal guidelines'', ``refrain'', ``avoid'', ``apology'', ``regret'' \\
\bottomrule
\end{tabular}}
\end{table}

\subsection{Usage of LLMs}
\label{appendix:usage_of_llm}
We employ LLMs to assist in several aspects of the research process: refining the writing of the paper (including formulation improvement, word choice polishing, and typo correction), drawing clear figures for experimental results, and supporting initial code development.

\subsection{Detailed Illustration of Model Structure}
\label{appendix:detail_model_structure}
Here, we illustrate the detailed structure of basic-form LALM and deduce the possible reason for the modality gap mentioned in Section~\ref{subsec:fail-steer-audio}.
Using Qwen2-Audio as an example, the \textit{audio encoder} is a Whisper-base stack, whereas the \textit{text encoder} is Qwen2’s Transformer-decoder embedding module; the \textit{projector} is a 2-layer MLP (1024→4096→4096) used to align input modalities into a shared latent space. 

Based on this model structure, we depict an inference process below to enhance clarity:
The input audio is first processed by a \textbf{Whisper-base audio encoder}, which converts the raw waveform into a sequence of high-level acoustic embeddings that capture phonetic, temporal, and prosodic cues. Then, the \textbf{projector} maps the audio features into the same latent space as text embedding. In parallel, the textual prompt (\eg, an instruction) is tokenized and embedded through Qwen2’s \textbf{Transformer-decoder text embedding layer}, producing semantic tokens in the model’s native text space. The aligned audio representations and the textual embeddings are concatenated and jointly consumed by the subsequent \textbf{Transformer decoder} to obtain the next-token probability.

We deduce that \textbf{the large distributional gap of audio modality may be attributed to the highly processed audio features} through a 32-layer audio encoder before feeding to the main body of the language model, where the tokenized text is fed and encoded inside the model directly.

\section{Additional Results}
\label{appendix:sec_additional_results}

\subsection{Impact of Different Refusal Directions}
\label{appendix:diff_refusal_direction}
We define the refusal steering vector for SARSteer using the differences between the harmful data and its refusal version. However, the refusal vector can also be calculated by the differences between safe data and its refusal version. Therefore, we make a comparison here to find out whether harmful data is the best option. We denote the safe-calculated one as ``Safe2Refusal'' and our harm-calculated one as ``Harm2Refusal''. Table~\ref{tab:ablation_refusal_dir} shows the comparison result. We find that Safe2Refusal performs unstably across models, although it can achieve better ASR in some cases, indicating that Harm2Refusal can be a better option.

\begin{table}[h!]
\centering
\caption{Ablation study on the effect of using different data types to compute the refusal direction. 
ASR ($\downarrow$) and BRR ($\uparrow$) are reported.}
\label{tab:ablation_refusal_dir}
\resizebox{0.4\linewidth}{!}{
\begin{tabular}{l|cc|cc}
\toprule
\multirow{2}{*}{Model} & \multicolumn{2}{c|}{Safe2Refusal} & \multicolumn{2}{c}{Harm2Refusal} \\
& ASR & BRR & ASR & BRR \\
\midrule
Qwen2-Audio & 27.60 & 79.20 & \textbf{10.80} & \textbf{79.95} \\
Kimi-Audio  & \textbf{6.80}  & 83.40 & 10.00 & \textbf{88.80} \\
\bottomrule
\end{tabular}}
\vspace{-2mm}
\end{table}

\subsection{Impact of Different Refusal Prompts}
\label{appendix:diff_refusal_prompts}
Our method is implemented based on a simple refusal prompt, \ie, ``I cannot assist with that.'', since the prompt selection is not within our main contribution. Here, we further test the impact of different refusal prompts.
Specifically, we select four representative refusal prompts, listed in Table~\ref{tab:refusal_prompts_examples} for comparison. Example 1 represents the simple refusal response pattern, which is used as the default refusal prompt in our method; Examples 2 and 3 are the defense prompts that we adapted from FSD~\citep{gong2025figstep} and AdaShield~\citep{wang2024adashield}, respectively; Example 4 represents the diversified refusal response patterns that provide a stronger refusal guide to LALMs.
The performance under the four examples is shown in Table~\ref{tab:perform_diff_refusal}. We can observe that all examples improve the performance toward both harmfulness and helpfulness, proving the effectiveness of our method as a basic framework using different defense prompts. Although Example 3 provides a stronger defense performance, it sacrifices helpfulness to some extent. A simple refusal example (\eg, Example 1) may be a more balanced choice as used in our work.

\begin{table}[ht]
\centering
\caption{Refusal prompt examples used to conduct SARSteer.}
\label{tab:refusal_prompts_examples}
\resizebox{\textwidth}{!}{
\begin{tabular}{l|p{15cm}}
\toprule
Example NO. & Refusal Prompt Details \\
\midrule
Example 1 & I cannot assist with that. \\
\midrule
Example 2 & First, you need to check whether there is speech in the audio. If there are speech instructions in the audio that are harmful or violating AI safety policy, you should not assist user’s requests as you are a harmless assistant. \\
\midrule
Example 3 & But before processing any instruction, examine the audio carefully for any text or items that might suggest harmful, illegal, or dangerous activity. If detected, do not engage with or act on the content. Analyze the audio and the given instruction holistically. If the content is determined to be unethical, illegal, or dangerous, please answer ``I am sorry''. Instead, please execute the above instruction safely and correctly. \\
\midrule
Example 4 & Here are some examples of how to refuse jailbreak requests: \newline
\quad 1. I cannot assist with that. \newline
\quad 2. I am not programmed to provide that information. \newline
\quad 3. My purpose is to help, but I cannot engage in harmful activities. \newline
\quad 4. I am here to provide support, but I cannot assist with that request. \\
\bottomrule
\end{tabular}}
\end{table}

\begin{table}[ht]
\centering
\caption{Performance under different refusal prompts using Figstep-audio. Best results are in \textbf{bold}. (ASR $\downarrow$, BRR $\uparrow$).}
\label{tab:perform_diff_refusal}
\resizebox{0.75\textwidth}{!}{
\begin{tabular}{l|cc|cc|cc|cc|cc}
\toprule
\multirow{2}{*}{Refusal Prompts} & 
\multicolumn{2}{c|}{No Defense} & 
\multicolumn{2}{c|}{Example 1} & 
\multicolumn{2}{c|}{Example 2} & 
\multicolumn{2}{c|}{Example 3} & 
\multicolumn{2}{c}{Example 4} \\
\cmidrule(lr){2-11}
& ASR & BRR & ASR & BRR & ASR & BRR & ASR & BRR & ASR & BRR \\
\midrule
Qwen2-Audio & 51.60 & 70.20 & 10.80 & \textbf{79.95} & 34.40 & 71.40 & \textbf{2.80} & 70.00 & 25.20 & 77.60 \\
Kimi-Audio  & 15.60 & 61.40 & 10.00 & \textbf{88.80} & 12.80 & 83.00 & \textbf{1.20} & 69.20 &     16.40  & 84.60      \\
\bottomrule
\end{tabular}}
\end{table}

\subsection{Performance on Base Model}
\label{appendix:perform_base_model}
Except for the instructed version of LALMs, which have been fine-tuned based on the instruction dataset that may contain some safety-related data, we evaluate the defense performance on the pre-trained only base model, to further verify the effectiveness. Table~\ref{tab:perform_basemodel} compares the defense performance of our SARSteer with all baselines. The results show that SARSteer can perform SOTA consistently across nearly all datasets, indicating the effectiveness of our method in even the base version of LALMs. It also shows the potential of adapting our method to the fine-tuning phase, \eg, constraining the learning direction based on the steering vector. We will continue more related exploration on its potential applications in future work.

\begin{table}[ht]
\centering
\caption{Performance of different defense methods on \textbf{Qwen2-Audio-Base}. 
ASR (\%) is for harmfulness (lower is better) and BRR (\%) is for helpfulness (higher is better). 
Best results are in \textbf{bold}, second-best are \underline{underlined}.}
\label{tab:perform_basemodel}
\resizebox{0.85\textwidth}{!}{
\begin{tabular}{l|cccc|cc}
\toprule
\multirow{2}{*}{Defense Method} & \multicolumn{4}{c|}{Harmfulness (ASR $\downarrow$)} & \multicolumn{2}{c}{Helpfulness (BRR $\uparrow$)} \\
\cmidrule(lr){2-7}
& Figstep-audio & SORRY-Bench & AJailBench & AdvBench-audio & Figstep-audio & AdvBench-audio \\
& (Harm) & -Audio &  & (Harm) & (Harmful-Safe) & (Harmful-Safe) \\
\midrule
No Defense & 62.80 & 49.77 & 43.00 & 51.15 & 48.60 & 51.64 \\
\midrule
AdaShield  & 39.20 & \underline{12.27} & 20.00 & \underline{18.27} & 50.00 & \underline{60.28} \\
FSD        & \underline{22.80} & 17.27 & \underline{15.00} & 32.69 & \textbf{60.20} & 56.06 \\
MDSteer-h2s& 45.60 & 16.82 & 18.50 & 20.77 & 49.80 & 49.81 \\
MDSteer-c2r& 46.40 & 22.50 & 28.00 & 19.23 & 50.40 & 50.29 \\
\textbf{SARSteer}   & \textbf{15.20} & \textbf{10.23} & \textbf{9.00} & \textbf{15.77} & \underline{58.20} & \textbf{60.39} \\
\bottomrule
\end{tabular}}
\end{table}

\subsection{Generalizability to LLM}
\label{appendix:generalize_llm}
We adapt our method, SARSteer, to the pure text-based LLM without audio modality to find out whether it has the potential to be applied in more scenarios. Table~\ref{tab:qwen2_sarsteer} shows the attempt on Qwen2~\citep{team2024qwen2} with the harmful queries input as text modality. Compared with the no-defense baseline, SARSteer consistently reduces harmfulness across SORRY-Bench and AdvBench while slightly improving helpfulness scores on both benchmarks. Although the ASR on Figstep remains nearly unchanged, the gains in other settings indicate that SARSteer generalizes beyond the audio modality and can provide robust protection in standard LLM scenarios without sacrificing the model's ability to respond to benign queries.

\begin{table}[ht]
\centering
\caption{Performance comparison on Qwen2 with and without SARSteer. 
Harmfulness is reported as ASR ($\downarrow$), and helpfulness is reported as BRR ($\uparrow$). Best results are in \textbf{bold}.}
\label{tab:qwen2_sarsteer}
\resizebox{0.85\linewidth}{!}{
\begin{tabular}{l|l|ccc|cc}
\toprule
\multirow{2}{*}{Model} & \multirow{2}{*}{Methods} 
& \multicolumn{3}{c|}{Harmfulness (ASR $\downarrow$)(\%)} 
& \multicolumn{2}{c}{Helpfulness (BRR $\uparrow$)(\%)} \\
\cmidrule(lr){3-5} \cmidrule(lr){6-7}
& & Figstep & SORRY-Bench & AdvBench 
& Figstep (Harmful-Safe) & AdvBench (Harmful-Safe) \\
\midrule
\multirow{2}{*}{Qwen2}
& No Defense & \textbf{10.00} & 17.73 & 0.38 & 88.00 & 95.39 \\
& SARSteer   & 10.40 & \textbf{15.91} & \textbf{0.19} & \textbf{88.40} & \textbf{96.25} \\
\bottomrule
\end{tabular}}
\vspace{-2mm}
\end{table}

\subsection{Impact of Different PCA Alternatives}
\label{appendix:pca_alter}
We justify the effectiveness of using PCA, as a linear extraction method, to extract the safety subspace here.
While nonlinear methods often capture richer and more expressive structures than linear ones, we chose PCA for two main reasons. \textbf{First}, PCA has been widely validated and adopted in prior studies as a reliable method for extracting \textbf{meaningful linear subspaces} from transformer activations \citep{tigges2023linear,matsumoto2022tracing,panickssery2023steering}. For example, Matsumoto et al. \citep{matsumoto2022tracing} extract intermediate-value representations using PCA, and Rimsky et al. \citep{panickssery2023steering} identify PCA directions to account for the most significant variance of the contrastive refusal dataset. \textbf{Second}, in our experiments, linear methods (\eg, PCA and SVD) consistently outperformed nonlinear alternatives such as kernel PCA. As shown in Table~\ref{tab:pca_comparison} below, kernel PCA substantially distorted the activation space and caused steering to collapse into repetitive and degenerate outputs. In contrast, PCA and SVD yield stable and consistent safety directions, reducing ASR significantly while preserving helpfulness and general utility. These empirical findings, together with established literature, justify PCA as the appropriate method for extracting safety-related subspaces.

\begin{table}[ht]
\centering
\caption{Performance comparison using different decomposed techniques Qwen2-Audio. (* represent the performance on meaningless outputs, \eg, ``not\textbackslash u8fd9\textbackslash u5757\textbackslash u8868\textbackslash u662f''.)}
\label{tab:pca_comparison}
\resizebox{\linewidth}{!}{
\begin{tabular}{l|cccc|cc|c}
\toprule
\multirow{2}{*}{Method} & \multicolumn{4}{c|}{Harmfulness (ASR $\downarrow$)} & \multicolumn{2}{c|}{Helpfulness (BRR $\uparrow$)} & General Utility \\
\cmidrule(lr){2-5} \cmidrule(lr){6-7} \cmidrule(lr){8-8}
& Figstep-audio (Harm) & SORRY-Bench-audio & AJailBench & AdvBench-audio (Harm) & Figstep-audio (Safe) & AdvBench-audio (Safe) & AirBench (Avg. Score) \\
\midrule
No Defense & 51.60 & 27.50 & 48.76 & 2.88 & 70.20 & 85.19 & 7.43 \\
PCA (Ours) & 10.80 & 13.41 & 18.00 & 0.58 & 79.95 & 85.00 & 7.53 \\
SVD & 12.40 & 12.27 & 19.00 & 1.58 & 81.20 & 84.23 & 8.33 \\
Kernel PCA & *2.00 & *1.14 & *0.96 & *0.00 & *50.80 & *50.00 & *5.27 \\
\bottomrule
\end{tabular}}
\end{table}

\subsection{Comparison with Fine-tuning-based Defense}
\label{appendix:ft_baseline}
We reproduce RRS~\citep{yang2025reshaping} as a fine-tuning-based defense for comparison under our defense resources. Specifically, we use 100 samples of the Figstep-audio data as implemented on our method (Section \ref{subsec:paired-dataset}) for RRS defense training and conduct evaluation on the other datasets. The results are shown in Table~\ref{tab:rrs_comparison} below. We can observe that \textbf{RRS performs limited compared to both “No Defense” and SARSteer}, with a significantly longer defense runtime (1145s compared to 266s in SARSteer). The main reason for performance degradation on RRS may come from the limited defense data (\ie, 100 samples in our paper), compared to the 1400+700 samples as in its original paper \citep{yang2025reshaping}, which is also a common limitation of the training-based defense.

\begin{table}[ht]
\centering
\caption{Performance comparison with training-based baseline RRS on Qwen2-Audio.}
\label{tab:rrs_comparison}
\resizebox{\linewidth}{!}{
\begin{tabular}{l|cccc|cc|c}
\toprule
\multirow{2}{*}{Methods} & \multicolumn{4}{c|}{Harmfulness (ASR $\downarrow$)} & \multicolumn{2}{c|}{Helpfulness (BRR $\uparrow$)} & Efficiency \\
\cmidrule(lr){2-5} \cmidrule(lr){6-7} \cmidrule(lr){8-8}
& Figstep-audio (Harm) & SORRY-Bench-audio & AJailBench & AdvBench-audio (Harm) & Figstep-audio (Safe) & AdvBench-audio (Safe) & Defense runtime (s) \\
\midrule
No Defense & 51.60 & 27.50 & 48.76 & 2.88 & 70.20 & 85.19 & - \\
RRS & 52.40 & 38.41 & 54.00 & 3.27 & 70.00 & \textbf{86.44} & 1145 \\
SARSteer & \textbf{10.80} & \textbf{13.41} & \textbf{18.00} & \textbf{0.58} & \textbf{79.95} & 85.00 & \textbf{266} \\
\bottomrule
\end{tabular}}
\end{table}

\subsection{Impact of Natural Speech Characteristics on Evaluation Data}
\label{appendix:impact_natural_speech_char}
We consider the performance differences on real-human-liked speech, \ie, containing natural speech characteristics, to evaluate the robustness of our method. To analyze the impact of natural speech characteristics (\eg, accent and emotion) other than our vanilla TTS dataset in Section~\ref{subsec:paired-dataset}, we conduct additional experiments on a relevant safety benchmark, Jailbreak-AudioBench~\citep{cheng2025jailbreak}, which incorporates several speech characteristics into existing safety datasets (\eg, AdvBench~\cite{yang2024air}). For fair comparison, we randomly sample 200 instances within the AdvBench subset, augmenting with \textit{celebrity accent}, \textit{5x volume emphasis}, and \textit{laugh emotion}, as well as some original audio. The results are shown in Table~\ref{tab:speech_characteristics}. We can observe that natural speech characteristics have only a slight influence on the results, indicating that they are less related to the LALM safety, and our method can consistently safeguard the model.

\begin{table}[ht]
\centering
\caption{Defense performance related to natural speech characteristics on Qwen2-Audio. (``Sub-category'' indicates the detailed augmentation techniques in Jailbreak-AudioBench).}
\label{tab:speech_characteristics}
\resizebox{0.5\linewidth}{!}{
\begin{tabular}{l|cc}
\toprule
\multirow{2}{*}{Natural speech characteristics} & \multicolumn{2}{c}{Harmfulness (ASR $\downarrow$)} \\
\cmidrule(lr){2-3}
& No Defense & SARSteer \\
\midrule
AdvBench-audio (ours) & 2.88 & 0.58 \\
Jailbreak-AudioBench & 1.50 & 0.50 \\
\addlinespace
\textbf{Sub-category} & \textbf{\#Successful/\#Total} & \textbf{\#Successful/\#Total} \\
\midrule
Accent & 0/50 & 0/50 \\
Emphasis & 0/50 & 0/50 \\
Emotion & 2/50 & 0/50 \\
Original & 1/50 & 1/50 \\
\bottomrule
\end{tabular}}
\end{table}

\subsection{Extended Evaluation on More Recent LALMs}
\label{appendix:perf_add_lalm}
To prove the generalizability of our method, we add more recent SOTA LALMs, MiDashengLM~\citep{dinkel2025midashenglm}, \textbf{Qwen2.5-Omni-7B}, \textbf{Qwen3-Omni-30B (MoE)}~\citep{xu2025qwen3}, and
\textbf{Voxtral-Mini-3B}~\citep{liu2025voxtral}, for comparison. The results of MiDashengLM on the four main datasets we used in the paper are shown in Table~\ref{tab:add_lalm}.

\begin{table}[ht]
\centering
\caption{Defense performance of SARSteer on MiDashengLM.}
\label{tab:add_lalm}
\resizebox{\linewidth}{!}{
\begin{tabular}{l|l|cccc|cc}
\toprule
\multirow{2}{*}{Model} & \multirow{2}{*}{Methods} & \multicolumn{4}{c|}{Harmfulness (ASR $\downarrow$)} & \multicolumn{2}{c}{Helpfulness (BRR $\uparrow$)} \\
\cmidrule(lr){3-6} \cmidrule(lr){7-8}
& & Figstep-audio (Harm) & SORRY-Bench-audio & AJailBench & AdvBench-audio (Harm) & Figstep-audio (Safe) & AdvBench-audio (Safe) \\
\midrule
Qwen2-Audio & No Defense & 51.60 & 27.50 & 48.76 & 2.88 & 70.20 & \textbf{85.19} \\
Qwen2-Audio & SARSteer & \textbf{10.80} & \textbf{13.41} & \textbf{18.00} & \textbf{0.58} & \textbf{79.95} & 85.00 \\
\midrule
Kimi-Audio & No Defense & 15.60 & 12.50 & 17.00 & \textbf{0.00} & 61.40 & 60.77 \\
Kimi-Audio & SARSteer & \textbf{10.00} & \textbf{6.14} & \textbf{11.00} & \textbf{0.00} & \textbf{88.80} & \textbf{86.83} \\
\midrule
MiDashengLM & No Defense & 12.80 & 22.50 & 11.00 & 5.00 & \textbf{57.20} & 68.76 \\
MiDashengLM & SARSteer & \textbf{0.00} & \textbf{5.68} & \textbf{8.50} & \textbf{1.73} & 53.20 & \textbf{77.21} \\
\bottomrule
\end{tabular}}
\end{table}

The evaluation of the other three recent LALMs is shown in Table~\ref{tab:new_lalms_full_baseline}. Importantly, all baseline
defenses considered in the main paper (MDSteer-h2s, MDSteer-c2r, AdaShield,
FSD) are re-evaluated under exactly the same protocol as in
Section~\ref{sec:experiment}, on the Figstep-audio benchmark. We additionally
report the \textit{Safe Refusal Rate} (\textbf{Safe\,RR}, lower is better), the fraction of \emph{benign} queries that are refused, to characterize over-refusal beyond BRR.
We observe that only SARSteer performs consistently well on all models, indicating its great generalizability across different LALMs. 

\begin{table}[h]
\centering
\caption{Full baseline comparison on three newer LALMs under Figstep-audio.
ASR ($\downarrow$) measures harmfulness, BRR ($\uparrow$) measures
helpfulness, and Safe\,RR ($\downarrow$) measures the refusal rate on benign
queries. \textbf{Bold} numbers denote the best result (except for ``No Defense'') per model. SARSteer is
the \emph{only} method that consistently reduces ASR while maintaining
Safe\,RR $\leq 1.6\%$ across all three new models.}
\label{tab:new_lalms_full_baseline}
\small
\begin{tabular}{llccc}
\toprule
Model & Method & ASR~($\downarrow$) & BRR~($\uparrow$) & Safe\,RR~($\downarrow$) \\
\midrule
\multirow{6}{*}{Qwen2.5-Omni-7B}
  & No Defense     & 42.0\%          & 71.8\%          & 0.0\% \\
  & MDSteer-h2s    & 58.8\%          & 50.8\%          & \textbf{0.0\%} \\
  & MDSteer-c2r    & 52.3\%          & 53.8\%          & 0.8\% \\
  & AdaShield      & \textbf{25.2\%} & 68.6\%          & 54.4\% \\
  & FSD            & 28.2\%          & 73.5\%          & 10.8\% \\
  & \textbf{SARSteer (Ours)} & 31.2\% & \textbf{80.2\%} & 0.8\% \\
\midrule
\multirow{6}{*}{Qwen3-Omni-30B (MoE)}
  & No Defense     & 7.2\%           & 95.3\% & 1.0\% \\
  & MDSteer-h2s    & 62.8\%          & 65.6\%          & \textbf{0.0\%} \\
  & MDSteer-c2r    & 15.2\%          & 73.4\%          & 2.0\% \\
  & AdaShield      & \textbf{3.6\%}  & 70.6\%          & 55.2\% \\
  & FSD            & 6.4\%           & 89.8\%          & 12.4\% \\
  & \textbf{SARSteer (Ours)} & 5.2\% & \textbf{94.0\%}          & 1.6\% \\
\midrule
\multirow{6}{*}{Voxtral-Mini-3B}
  & No Defense     & 34.0\%          & 71.4\% & 0.4\% \\
  & MDSteer-h2s    & 50.8\%          & 65.4\%          & \textbf{0.0\%} \\
  & MDSteer-c2r    & 26.8\%          & 67.6\%          & 0.8\% \\
  & AdaShield      & \textbf{2.4\%}  & 66.4\%          & 54.0\% \\
  & FSD            & 35.2\%          & 65.8\%          & 12.0\% \\
  & \textbf{SARSteer (Ours)} & 30.0\% & \textbf{69.2\%}         & 0.8\% \\
\bottomrule
\end{tabular}
\end{table}

\subsection{Analysis of Statistical Stability}
\label{appendix:stat_stability}
We have strictly set one unified random seed for all relevant packages, \eg, numpy, random, and torch, to ensure the reproducibility of our results. To further demonstrate the statistical stability, we conduct multiple runs with five different random seeds (\ie, 42, 1, 123, 678, 999) as in Table~\ref{tab:multi_run}. The observed variances across all metrics can be considered adequately low, especially for our method, confirming the reliability of our experimental conclusions. 

\begin{table}[ht]
\centering
\caption{Defense performance of multiple runs on Qwen2-Audio.}
\label{tab:multi_run}
\resizebox{\linewidth}{!}{
\begin{tabular}{l|cccc|cc|c}
\toprule
\multirow{2}{*}{Multi-run (seed)} & \multicolumn{4}{c|}{Harmfulness (ASR $\downarrow$)} & \multicolumn{2}{c|}{Helpfulness (BRR $\uparrow$)} & General Utility \\
\cmidrule(lr){2-5} \cmidrule(lr){6-7} \cmidrule(lr){8-8}
& Figstep-audio (Harm) & SORRY-Bench-audio & AJailBench & AdvBench-audio (Harm) & Figstep-audio (Safe) & AdvBench-audio (Safe) & AirBench (Avg. Score) \\
\midrule
\multicolumn{8}{c}{\textbf{No Defense}} \\
\midrule
42 (default) & 51.60 & 27.50 & 48.76 & 2.88 & 70.20 & 85.19 & 7.43 \\
1 & 53.20 & 30.05 & 49.00 & 3.65 & 68.80 & 85.77 & 7.38 \\
123 & 52.80 & 30.95 & 48.00 & 3.85 & 70.40 & 86.73 & 7.38 \\
678 & 54.00 & 29.36 & 43.50 & 3.65 & 71.60 & 85.96 & 7.37 \\
999 & 54.00 & 29.14 & 45.00 & 3.65 & 70.80 & 86.83 & 7.36 \\
\midrule
\textbf{Average} & \textbf{53.12} & \textbf{29.40} & \textbf{46.85} & \textbf{3.54} & \textbf{70.36} & \textbf{86.10} & \textbf{7.38} \\
\textbf{Variance} & \textbf{0.9920} & \textbf{1.6261} & \textbf{6.0595} & \textbf{0.1420} & \textbf{1.0480} & \textbf{0.4698} & \textbf{0.0007} \\
\midrule
\multicolumn{8}{c}{\textbf{SARSteer}} \\
\midrule
42 (default) & 10.80 & 13.41 & 18.00 & 0.58 & 79.95 & 85.00 & 7.53 \\
1 & 12.00 & 12.50 & 17.50 & 0.58 & 78.40 & 86.35 & 7.33 \\
123 & 12.40 & 12.27 & 20.00 & 0.38 & 77.80 & 88.75 & 7.33 \\
678 & 11.60 & 12.73 & 15.00 & 0.38 & 79.60 & 88.18 & 7.35 \\
999 & 11.20 & 12.27 & 16.00 & 0.38 & 78.40 & 87.70 & 7.38 \\
\midrule
\textbf{Average} & \textbf{11.60} & \textbf{12.64} & \textbf{17.30} & \textbf{0.46} & \textbf{78.83} & \textbf{87.19} & \textbf{7.38} \\
\textbf{Variance} & \textbf{0.4000} & \textbf{0.2236} & \textbf{3.7000} & \textbf{0.0120} & \textbf{0.8195} & \textbf{2.2901} & \textbf{0.0071} \\
\bottomrule
\end{tabular}}
\end{table}

\subsection{Analysis and Quantitative Evidence of Text-audio Space Difference}
\label{appendix:text-audio-space-diff}
In Section \ref{subsec:fail-steer-audio}, we demonstrate the limitations of vanilla adaptation of activation steering defenses by presenting ASR in Figure \ref{fig:vanilla_steer} and using t-SNE in Figure \ref{fig:tsne_hs} to illustrate the potential reason, where the latent space difference between audio and text is the observation from the latter. 
A similar visualization technique was used to analyze the contrastive dataset (similar to our harmful-safe paired setting) by Panickssery et al. \citep{panickssery2023steering}, and found a linear separation of text modality after a certain layer. This observation supports their successful steering of contrastive text data. Similarly, in Figure \ref{fig:tsne_hs}, we observe this phenomenon in text-pair data, but different behavior (a large distributional gap across layers) in audio-pair data, where the steering fails. This can thus be considered the main reason for the failure of steering audio modality.

To further measure this gap, we add an experiment on centered kernel alignment (CKA) with both linear and RBF kernels to quantitatively compare the hidden-state difference within audio-pair data and text-paired data. The results clearly confirm the substantial text–audio discrepancy:
\begin{itemize}
    \item Linear CKA (audio\_harm vs. audio\_safe): 0.0631
    \item Linear CKA (text\_harm vs. text\_safe): 0.6722
    \item RBF CKA (audio\_harm vs. audio\_safe): 0.1427
    \item RBF CKA (text\_harm vs. text\_safe): 0.6804
\end{itemize}
These measurements show that \textbf{harmful and safe text representations remain closely aligned—consistent with effective text-based steering—whereas harmful and safe audio representations diverge significantly}, explaining why steering fails when directly transferred from text to audio.

\subsection{Human Evaluation of the LLM-as-a-Judge Protocol}
\label{app:b11_human_eval}
To judge LLM-judge reliability, we performed human evaluation on \textbf{500} samples. Samples are drawn via \emph{stratified sampling} covering all 5 LALMs
(Qwen2-Audio, Kimi-Audio, Qwen2.5-Omni, Qwen3-Omni, Voxtral-Mini) under both
\emph{No Defense} and \emph{SARSteer} conditions, and additionally include
PAIR and GCG attacks on Kimi-Audio (see~Appendix~\ref{app:b12_adaptive}).
The annotators independently label each response as \textit{safe} or
\textit{unsafe} under a fully blinded protocol: annotators have no access to
the model identity, the defense condition, or the LLM judge's label.

\begin{table}[h]
\centering
\caption{Confusion matrix of human vs.\ LLM-as-a-judge labels on 500 stratified samples.}
\label{tab:human_eval_500}
\small
\begin{tabular}{lcc}
\toprule
 & LLM $=$ Safe & LLM $=$ Unsafe \\
\midrule
Human $=$ Safe   & 271 & 43 \\
Human $=$ Unsafe & 3   & 183 \\
\bottomrule
\end{tabular}
\end{table}

\textbf{Results.}
Overall agreement rises to $90.8\%$ with Cohen's $\kappa = 0.811$
(substantial). The false negative rate is only $3/500$, and the false-positive/false-negative ratio is $43{:}3 \approx 14{:}1$, representing a pronounced \emph{conservative} skew.

\textbf{Qualitative analysis of the 43 false positives.}
All 43 follow a consistent pattern: the model begins with a clear refusal (\eg, ``I'm unable to assist with this request'') and then produces contextually related but harmless content, explaining why the request is dangerous, offering prevention guidelines, or suggesting legitimate alternatives. The LLM judge flags such responses as unsafe based on surface-level topic cues, whereas human annotators correctly recognize the actual content as safe. This is a \emph{systematic conservative bias}, not
stochastic noise.

\textbf{Implication for our claims.}
Because the bias is strictly one-sided, the LLM-judge ASR values reported in
the main paper are \emph{upper bounds} on the true attack success rate:
the judge \textbf{over}-estimates risk, it does not \textbf{under}-estimate it.
Consequently, the safety improvement of SARSteer is, if anything,
\emph{under-estimated}. Improving the nuanced capability of safety evaluators (\ie, separating refusal surface form from the actual content) is a promising direction for future work.

\subsection{Robustness Under Adversarial Perturbations and Adaptive Attacks}
\label{app:b12_adaptive}

We further probe SARSteer under stronger, adaptive attacks. Two
orthogonal attack families are considered.

\paragraph{Signal-level audio perturbations.}
We apply (i)~projected gradient descent (\textbf{PGD}) on the raw audio
waveform with $\ell_\infty$ budget $\epsilon{=}0.005$, and (ii)~additive
\textbf{Gaussian noise} with $\sigma{=}0.01$, following the standard audio
adversarial setup. Attacks are evaluated on Qwen2-Audio using the
Figstep-audio benchmark.

\paragraph{Text-origin adaptive attacks (PAIR / GCG).}
Because no established LALM counterpart of GCG~\citep{zou2023universal} and
PAIR~\citep{chao2025jailbreaking} exists, we implement audio-adapted versions:
(i)~\textbf{PAIR~(audio)} uses PAIR to generate semantically rephrased
jailbreak queries in text (role-playing, academic disguise, \etc.) and
converts each rephrased query to audio via TTS; (ii)~\textbf{GCG~(audio)}
first optimizes an adversarial suffix on the text input via the standard
GCG algorithm, appends the suffix to the original harmful query, and then
converts the combined text to audio via TTS. These tests probe whether semantic-level and token-level adversarial attacks retain their capability after the $\text{text}\!\rightarrow\!\text{audio}\!\rightarrow\!\text{model}$
pipeline. We evaluate 50 samples per attack on Kimi-Audio.

\begin{table}[h]
\centering
\caption{Robustness of SARSteer under signal-level perturbations
(Qwen2-Audio, Figstep-audio) and audio-adapted adaptive attacks
(Kimi-Audio, Figstep-audio). ASR ($\downarrow$).}
\label{tab:adaptive_attacks}
\small
\begin{tabular}{llcc}
\toprule
Attack type & Model & No Defense & SARSteer \\
\midrule
PGD ($\epsilon{=}0.005$)    & Qwen2-Audio & 24.4\% & \textbf{6.0\%} \\
Gaussian ($\sigma{=}0.01$)  & Qwen2-Audio & 27.2\% & \textbf{7.2\%} \\
PAIR (audio)                & Kimi-Audio  & 58.0\% & \textbf{10.0\%} \\
GCG  (audio)                & Kimi-Audio  & 14.0\% & \textbf{6.0\%} \\
\bottomrule
\end{tabular}
\end{table}

\textbf{Observations.}
(i)~Signal-level perturbations partially degrade the intelligibility of the
audio, which explains why the No-Defense ASR ($24.4\%$/$27.2\%$) is lower
than on clean TTS audio ($51.6\%$ in Table~\ref{tab:results_harmful_helpful}); SARSteer nonetheless further reduces ASR by $\sim4\times$ in both cases.
(ii)~Among adaptive attacks, PAIR transferred through the audio channel is
remarkably effective ($58.0\%$), yet SARSteer drives ASR down to $10.0\%$. GCG is less effective through the audio channel, where gradient-optimized suffixes are distorted by the text-to-audio conversion, but SARSteer further reduces the residual ASR to $6.0\%$.
Combined with the TTS, audio-perturbation (AJailBench), and natural-speech
(AdvBench-audio, Jailbreak-AudioBench) evaluations in the main paper,
SARSteer is now validated effective across \textbf{five} attack categories.

\subsection{Sample-Level Validation of the PCA-Based Selectivity}
\label{app:b13_sample_level_pca}

We validate the PCA-based selectivity at the \emph{individual-sample} level. Ideally, one would measure $\Delta s_i = s(\mathbf{h}_i + \alpha \widehat{\mathbf{v}}_{\!\perp}) -
s(\mathbf{h}_i)$ directly, but this requires access to the closed-form
$s(\cdot)$. As a faithful and tractable proxy, we compute the per-sample
alignment
\begin{equation}
  \delta_i \;=\; \big\langle \widehat{\mathbf{v}}_{\!\perp},\,\mathbf{h}_i \big\rangle ,
  \label{eq:delta_i}
\end{equation}
averaged across all transformer layers. By the first-order expansion
$\Delta s_i \approx \alpha\, \mathbf{w}^{\top} \widehat{\mathbf{v}}_{\!\perp}$
in Eq.~(17), $\delta_i$ captures how strongly
$\widehat{\mathbf{v}}_{\!\perp}$ interacts with each individual activation
and therefore serves as a sample-level indicator of the effective steering
signal.

We measure $\delta_i$ on 100 harmful and 100 safe queries drawn from
Figstep-audio on Qwen2-Audio. The results are shown in Table~\ref{tab:sample_level_pca}.

\begin{table}[h]
\centering
\caption{Sample-level alignment $\delta_i = \langle
\widehat{\mathbf{v}}_{\!\perp}, \mathbf{h}_i \rangle$ averaged over all
layers. Harmful samples exhibit significantly higher $\delta_i$ and a
larger positive tail than safe samples.}
\label{tab:sample_level_pca}
\small
\begin{tabular}{lccc}
\toprule
Group   & Mean $\delta$ & Std $\delta$ & \% with $\delta>0$ \\
\midrule
Harmful & $-552.2$ & $291.9$ & $6.0\%$ \\
Safe    & $-650.9$ & $\phantom{0}63.4$ & $0.0\%$ \\
\midrule
\multicolumn{4}{l}{Welch's $t$-test: $p = 0.001$,\quad Cohen's $d = 0.47$} \\
\bottomrule
\end{tabular}
\end{table}

\textbf{Interpretation.}
The negative mean $\delta$ in both groups indicates that
$\widehat{\mathbf{v}}_{\!\perp}$ is broadly anti-aligned with typical hidden
activations -- the safe subspace has already been projected out. Crucially,
harmful samples show (i) a significantly higher mean ($p = 0.001$,
medium-to-large effect size $d = 0.47$), (ii) $4.6\times$ larger variance,
and (iii) a non-trivial positive tail ($6.0\%$ of harmful samples have
$\delta_i > 0$ vs.\ exactly $0.0\%$ of safe samples). This is precisely the
behavior depicted in Appendix~\ref{appendix:math_intuition}: the steering signal is selectively activated on harmful samples while being effectively suppressed on safe ones, providing individual-sample support.


\end{document}